\documentclass[a4paper,11pt]{article}

\pdfoutput=1
\usepackage{comment}
\usepackage{fixltx2e}
\usepackage[table,xcdraw]{xcolor}
\usepackage{jcappub} 
\usepackage{float}
  
\definecolor{RoyalBlue}{rgb}{0.25,.41,.88}

\def\be{\begin{equation}}
\def\ee{\end{equation}}
\def\bea{\begin{eqnarray}}
\def\eea{\end{eqnarray}}

\def \bk {\mathbf{k}}

\usepackage[normalem]{ulem}

 \newcommand\q{{\bf q}}
 \renewcommand\k{{\bf k}}

 \newcommand{\n}{\nonumber \\}

\usepackage[T1]{fontenc}
\usepackage{graphbox}
\usepackage{subfig}
\usepackage{booktabs}
\subheader{TUM-HEP-1457/23}
\title{\boldmath The Effective Field Theory of Large-Scale Structure and Multi-tracer II: redshift space and realistic tracers}
\usepackage{amsmath}
\author[a]{Thiago Mergulh\~ao,}
\author[b]{Henrique Rubira,}
\author[c]{Rodrigo Voivodic}

\affiliation[a]{\small Institute for Astronomy, University of Edinburgh, Royal Observatory, Blackford Hill, Edinburgh EH9 3HJ, UK}
\affiliation[b]{\small Physik Department T31, Technische Universit\"at M\"unchen,\\
James-Franck-Stra{\ss}e 1, D-85748 Garching, Germany}
\affiliation[c]{\small Donostia International Physics Center (DIPC), Paseo Manuel de Lardizabal, 4, 20018 Donostia-San Sebasti\'an, Spain}

\emailAdd{thiago.mergulhao@ed.ac.uk}
\emailAdd{henrique.rubira@tum.de}
\emailAdd{rodrigo.voivodic@gmail.com}

\abstract{We extend the multi-tracer (MT) formalism of the effective field theory of large-scale structure to redshift space, comparing the results of MT to a single-tracer analysis when extracting cosmological parameters from simulations. We used a sub-halo abundance matching method to obtain more realistic multi-tracer galaxy catalogs constructed from N-body simulations. Considering different values for the sample shot noise and volume, we show that the MT error bars on $A_s$, $\omega_{\rm cdm}$, and $h$ in a full-shape analysis are approximately $50\%$ smaller relative to ST. We find that cosmological and bias coefficients from MT are less degenerate, indicating that the MT parameter basis is more orthogonal. We conclude that using MT combined with perturbation theory is a robust and competitive way to accommodate the information present in the mildly non-linear scales.
}

\keywords{Large Scale Structure, Power Spectrum, Multi-tracer, Perturbation Theory}

\begin{document}

\maketitle
\flushbottom

\section{Introduction}
\label{sec:introduction}
The large-scale structure of the Universe (LSS) is a key probe for cosmology. Ongoing spectroscopic and photometric surveys, such as BOSS, eBOSS, and DES  \cite{BOSS:2016wmc,Abbetal,SDSS-IV:2019txh} have enlarged our knowledge of structure formation and in the short future DESI, Euclid, and LSST \cite{DESI:2016fyo,Amendola:2012ys,Ivezic:2008fe} data will also sum up to current analysis. LSS data will be vital, for instance, to shed light on the tensions in the values of the Hubble constant \cite{Schoneberg:2021qvd} and the amplitude of primordial fluctuations. 

The current state-of-the-art LSS analysis includes extracting data from the (mildly) non-linear regime using the full-shape (FS) modeling of the galaxy power spectrum and higher-order n-point functions. This approach builds upon the effective field theory of LSS (EFTofLSS) \cite{Baumann:2010tm, Carrasco:2012cv, Carrasco:2013mua, Konstandin:2019bay,Angulo:2015eqa,Baldauf:2021zlt} and the large-scale bias expansion of matter tracers \cite{Assassi2014,Desjacques2016}. It has achieved remarkable results when applied to BOSS data \cite{Sanchez:2013tga,DAmico:2019fhj,Ivanov:2019hqk,Colas:2019ret, Philcox:2020vvt, Nishimichi:2020tvu, Semenaite:2021aen} and when probing new physics \cite{Ivanov:2020ril,Lague:2021frh, Semenaite:2022unt,Rubira:2022xhb, Piga:2022mge,Carrilho:2022mon, Simon:2022ftd,Herold:2021ksg}. One of its downsides is that it requires a few free parameters (counter-terms, bias, and stochastic coefficients) to be fitted in data, although those parameters have a clear physical motivation and guarantee the correct parametrization of structure formation on the mildly non-linear regime. The quality of the EFT modeling is therefore attached to how well those free parameters can be determined by the data available. 

In parallel, multi-tracer (MT) techniques are another way to boost the information extracted from galaxy surveys and have been used to probe primordial non-gaussianities and redshift-space distortions \cite{Seljak:2008xr, McDonald:2008sh, Abramo2013, Abramo:2015iga, Abramo:2022qir, Barreira:2023rxn,Karagiannis:2023lsj,Blake:2013nif,Ross:2013vla,Beutler:2015tla,Marin:2015ula,Zhang:2021uyp,Sullivan:2023qjr}. MT was also developed in the context of the eBOSS, both in configuration and Fourier space, considering correlations between luminous red galaxy (LRG) and emission line galaxy (ELG) \cite{Wang:2020tje,Zhao:2020tis} and also voids \cite{Zhao:2021ahg}. In those works, the principal motivation for using MT is that it leads to cosmic variance cancellation.

As we have motivated in \cite{Mergulhao:2021kip}, the gains of doing multi-tracer go beyond cosmic variance cancellation in linear scales. 
We have shown that a real-space FS analysis of MT combined with EFTofLSS outperforms the usual single-tracer (ST) approach, improving the constraints on cosmological parameters with error bars approximately 60\% smaller relative to ST. The MT improvement can be attributed to a degeneracy break between cosmological and EFT parameters. Via the MT splitting of a tracer into sub-samples, one can single out non-linear effects in the gravitational evolution of each sub-species and add the cross power spectrum as a new source of information. On the other hand, single-tracer analysis smooths out non-linear dynamics for each tracer into a single averaged specie. Therefore, despite MT having more free parameters, we concluded that those parameters were {\it better determined} by the same data. 

In this work, we extend the EFTofLSS results in redshift space to allow for multiple tracers and test it by performing a set of Markov chain Monte Carlo (MCMC) parameter extraction from simulations. We use the \texttt{TheOne} simulation from the \texttt{BACCO} project \cite{Angulo:2020vky, Zennaro:2021bwy} together with the {\it subhalo abundance matching extended} (SHAMe) method to populate halos with galaxies \cite{Contreras:2020him}. We also perform a tracer split into two populations following the galaxy's star formation rate (SFR) provided by SHAMe, which is very similar to dividing galaxies by their color \cite{SFRcolor}. This improvement makes the comparison to galaxy surveys more reliable compared to our previous analysis \cite{Mergulhao:2021kip}, which focused mainly on halos split by their masses. Moreover, we analyze the dependence of the MT method on the shot noise of the sample and the volume of the box considered, providing projections for MT performance for ongoing and future galaxy surveys.
We find a substantial improvement in the extraction of cosmological parameters (and also in the bias parameters), ranging from $40\%$ to $80\%$ depending on the scales and samples considered. Our results show that MT method strongly breaks the degeneracy between cosmological and the EFT parameters, indicating that there is considerable information gain in the tracer splitting.

This paper is structured as follows: we described the theoretical model of MT in redshift space in Sec.~\ref{sec:theory}. The simulation data, the covariance, and the MCMC setup used are described in Sec.~\ref{sec:data}. We present the main results in Sec.~\ref{sec:results} and conclude in Sec.~\ref{sec:conclusion}. 

\section{Theoretical model}
\label{sec:theory}

In this section, we review the FS modeling of tracers in both real and redshift space for a single-tracer. 
We then generalize it to accommodate multiple tracers.

\subsection{Single-tracer (ST)}

\paragraph{Real space}

The most general bias expansion for the overdensity of a tracer $A$, $\delta^A$, in an operator basis $\mathcal{O}$ is given by \cite{Assassi2014,Desjacques2016}
\begin{equation}
\label{eqn: bias_expansion}
    \delta^{A}(\boldsymbol{x}, \tau)=\sum_{\mathcal{O}} b_{\mathcal{O}}(\tau) \mathcal{O}(\boldsymbol{x}, \tau) + \epsilon(\boldsymbol{x}, \tau) + \sum_\mathcal{O}\epsilon_\mathcal{O}(\boldsymbol{x}, \tau)\mathcal{O}(\boldsymbol{x}, \tau)\,,
\end{equation}
where $\epsilon (\boldsymbol{x}, \tau)$and $\epsilon _{\mathcal{O}} (\boldsymbol{x}, \tau)$ are stochastic fields and $b_\mathcal{O} (\tau)$ are the tracer (time-dependent) bias coefficients. Up to the third order in the fields, we find the following set of relevant operators
\begin{equation}
\label{eqn: operator_basis}
    \mathcal{O} \in
    \left\{
    \delta, \delta^{2}, \delta^{3}, \mathcal{G}_{2}[\Phi_{g}], \delta\mathcal{G}_{2}[\Phi_{g}] , \mathcal{G}_{3}[\Phi_{g}], \Gamma_{3}[\Phi_{g}, \Phi_{v}], \nabla^2 \delta \right\}\,,
\end{equation}
with $\delta$ being the matter overdensity, $\mathcal{G}_2$ an $\mathcal{G}_3$ being the second and third-order Galileon operators and $\Gamma_3$ a combination of Galileon operators for the gravitational $\Phi_{g}$ and velocity potential $\Phi_{v}$. The last term in the set above, $\nabla^2 \delta$, accounts for the fact that tracer formation occurs over a finite region \cite{Desjacques2016,Lazeyras:2019dcx} and it will lead to terms proportional to $k^2 \delta$.

Using the set of operators in Eq.~(\ref{eqn: bias_expansion}), the auto power-spectrum of tracer $A$ can be computed as \cite{Assassi2014}
\begin{eqnarray} 
\label{eq:generalP}
 P^{AA}(z,k) &=& \left[b^A_1(z)\right]^2 \left[ P_{\rm lin}(k) + P_{\rm 1L}(k) \right] + b^A_1(z)b^A_2(z)\,\mathcal{I}_{\delta^2}(k) + 2b^A_1(z)b^A_{\mathcal{G}_2}(z)\,\mathcal{I}_{\mathcal{G}_2}(k) \nonumber \\
 &+& \left[2b^A_1(z)b^A_{\mathcal{G}_2}(z) + \frac{4}{5} b^{A}_1(z)b^A_{\Gamma _{3}}(z)\right]\mathcal{F}_{\mathcal{G}_2}(k)  
+ \frac{1}{4}\left[b^A_2(z)\right]^2\,\mathcal{I}_{\delta^2\delta^2}(k)  \\ &+&\left[b^A_{\mathcal{G}_2}(z)\right]^2\,\mathcal{I}_{\mathcal{G}_2\mathcal{G}_2}(k) + b^A_2(z)b^A_{\mathcal{G}_2}(z)\,\mathcal{I}_{\delta^2\mathcal{G}_2}(k) + P_{\rm ct}^{AA}(k) + P_{\varepsilon^A\varepsilon^A}(k)  \,, \nonumber
\end{eqnarray}
where $P_{\rm lin}$ and $P_{\rm 1L}$ are respectively the linear matter spectrum and its one-loop correction. For a complete expression of the operators in Eq.~(\ref{eqn: operator_basis}) and the spectra in Eq.~(\ref{eq:generalP}), see \cite{chudaykin2020nonlinear}. Hereafter, we omit the time dependence. The $P_{\rm ct}$ term is a sum of two different contributions: the dark matter sound speed counter-term $c_s^2$ and the intrinsic halo-formation scale $R^A_*$
\begin{equation} \label{eq:nabla2}
    P_{\rm ct}^{AA}(k) = -2\left[\left(b^A_1\right)^2\frac{c_s^2}{k^2_{\mathrm{NL}}} + b^A_1\left(R^A_*\right)^2\right]k^2P_{\rm lin} \equiv -2b^A_{\nabla^2\delta}b^A_1\frac{k^2}{k_{\rm norm}^2}P_{\rm lin} \,.
\end{equation}
In the last step the two parameters $k_{\mathrm{NL}}$ and $1/R^A$, respectively the non-linear scale for matter and the typical halo scale, were rewritten as a single arbitrary scale, fixed to be $k_{\rm norm} = 0.1\,h\,\mathrm{Mpc}^{-1}$. We also redefined the bias parameter combination as $b^A_{\nabla^2\delta}$. 

Finally, the last term in Eq.~(\ref{eq:generalP}) is a stochastic contribution which can be written as an expansion in even powers of $k$ \cite{Desjacques2016}:
\begin{equation} \label{eq:shotnoise}
    P_{\varepsilon^A\varepsilon^A}(k) = \frac{1}{\Bar{n}_A}\left(1 + c^{AA}_{\rm st,0} + c^{AA}_{\rm st,2}\frac{k^2}{k_{\rm norm}^2} \right)\,
\end{equation}
with $\Bar{n}_A$ being the number density of the tracer $A$. Note that the stochastic parameters as defined above will capture deviations from perfect Poisson noise.

\paragraph{Redshift space}
We now move to calculate the structure of the auto-correlator for the tracer $A$ in redshift space, later expanding it in multipoles. After changing coordinates to redshift space, up to third-order in perturbations, the $AA$ spectrum is written as \cite{Perko2016}
\begin{eqnarray} \label{eq:rsd_st}
    P^{AA}(k,\mu)= & \left[Z^A_1(\k)\right]^2
P_{\rm lin}
(k)+ 2\int_{\q}\left[Z^A_2(\q,\k-\q)\right]^2
P_{\rm lin}(|\k-\q|)
P_{\rm lin}(q)+  \n
&  6Z_1^A(\k)P_{\rm lin}(k)\int_{\q}Z_3^A(\q,-\q,\k)P_{\rm lin}(q)+ P_{\text{ct}}^{AA}(k,\mu) + P_{\varepsilon^A\varepsilon^A}(k,\mu)\,,
\end{eqnarray}
in which the $Z$ kernels are described in \cite{Perko2016} and $\mu = \hat{z} \cdot \vec{k}/k$ is the angle w.r.t the line-of-sight direction $\hat{z}$. Following the parametrization of \cite{Perko2016}, the counter-terms are\footnote{Note that there are different parametrizations of the counter-terms and bias coefficients in redshift space \cite{Nishimichi:2020tvu}. In principle, those differences may lead to different constraints in cosmological parameters due to prior and volume effects \cite{Carrilho:2022mon,Simon:2022lde}. Those differences are however small in the limit in which the data set is constraining enough. }
\begin{eqnarray}\label{eq:ct_rsd}
   P^{AA}_{\mathrm{ct}}(k,\mu) &=& 2P_{11}Z^{\rm A}_1 \frac{k^2}{k_{\rm norm}^2} 
   \left[c^A_{{\rm ct},20} + c^A_{{\rm ct},22}\mu^2 + c^A_{{\rm ct},24}\mu^4 + c^A_{{\rm ct},26}\mu^6  \right. \nonumber \\ 
   && \left. +\, c^A_{{\rm ct},44} \frac{k^2}{k_{\rm norm}^2} \mu^4 + c^A_{{\rm ct},46}\frac{k^2}{k^2_{\rm norm}}\mu^6 \right] .
\end{eqnarray}
Similar to Eq.~(\ref{eq:nabla2}), we keep only a single normalization scale $k_{\rm norm}$. Notice that the redshift-space term $P_{\rm ct}(k,\mu)$ contains higher-order terms in $k^2$ when compared to its real-space version Eq.~(\ref{eq:nabla2}) and \cite{Perko2016}. Those terms serve as a proxy to capture higher-order contributions of fingers-of-god (FoG) \cite{Jackson:1971sky,Ivanov2019,Nishimichi:2020tvu}. We comment later on the importance of those terms for different tracers, highlighting the contribution of the $\mu^6$ term in the context of MT. 

The last term in Eq.~(\ref{eq:rsd_st}) comes from the stochastic contribution $\epsilon^{A}_{\rm RS}$, which also includes a velocity component
\begin{eqnarray}
  \epsilon^A_{\rm RS} = \delta^A_{\epsilon} + f\mu^2\theta^A_{\epsilon} \;,
\end{eqnarray}
where $\theta$ is the velocity divergence and $f$ is the log derivative of the growth factor $D$ with respect to the scale factor $f = d \log{D}/ d\log{a}$. Its power spectrum is given by\footnote{Regarding the $\theta^A_{\epsilon}$ contribution, due to momentum and mass conservation, we expect its first $k^0$ contribution to be the same as the matter one \cite{Desjacques2016}. The first (non-degenerate) contribution then should scale as $k^2$. Moreover, one should also consider stochastic contributions from the renormalized contact operators, but it happens that up to $k^2$ contributions, these are degenerate with the other $k^2$ contributions.}
\begin{equation}
    P_{\varepsilon^A\varepsilon^A}(k,\mu)  = \frac{1}{\bar{n}_A}\left[1+c^{AA}_{\mathrm{st},00} + c^{AA}_{\mathrm{st},20}\frac{k^2}{k_{\rm norm}^2} + c^{AA}_{\mathrm{st},22}\frac{k^2}{k_{\rm norm}^2}f\mu^2\right]\;.
\end{equation}
Therefore, a single-tracer analysis up to one-loop in perturbation theory in redshift space requires, in addition to the cosmological parameters, the fit of the following set of parameters:
{\small \begin{equation}  \label{eq:all_terms_ST}
    \left\{b^A_1,\, b^A_2,\, b^A_{\mathcal{G}_2},\, b^A_{\Gamma_3},\,c^A_{{\rm ct},20},\, c^A_{{\rm ct},22},\, c^A_{{\rm ct},24},\, c^A_{{\rm ct},26},\, c^A_{{\rm ct},44},\, c^A_{{\rm ct},46},\, c^{AA}_{\mathrm{st}, 00},\, c^{AA}_{\mathrm{st}, 20},\, c^{AA}_{\mathrm{st}, 22}\right\}\;.
\end{equation}}
We discuss the relevance of each term later in the text. 

We can now project the tracer spectrum into $\ell$-multipoles
\begin{eqnarray}
   P^{AA}(k,\mu) =\sum_{\ell \; {\rm even}} {\cal L}_\ell(\mu) P^{AA}_{\ell}(k)\,,
\end{eqnarray}
with
\begin{eqnarray} \label{eq:project}
   P^{AA}_{\ell}(k) \equiv \frac{2\ell+1}{2}\int_{-1}^{1}d\mu \, {\cal L}_\ell(\mu) 
 P^{AA} (k,\mu)\,,
\end{eqnarray}
to obtain the following map for the projected counter-terms onto the non-projected ones:
\begin{eqnarray}
P^{AA}_{\mathrm{ct, \;\ell = 0}}(k) &=& \frac{2}{105}\frac{k^2}{k_{\rm norm}^2}P_{\rm lin}(k)\left[35\left(f + 3 b^A_1\right)c^A_{\mathrm{ct},20} + 7\left( 3f + 5b^A_1\right)c^A_{\mathrm{ct}, 22}  \right.\\ \nonumber
 &+& \left. 3\left(5f + 7b^A_1\right)\left(c^A_{\mathrm{ct},24} + \frac{k^2}{k_{\rm norm}^2}c^{A}_{\mathrm{ct}, 44} \right) + \frac{35}{21}\left(7f + b^A_1\right)\left(c^A_{\mathrm{ct},26} + \frac{k^2}{k_{\rm norm}^2}c^A_{\mathrm{ct},46}\right)\right]\;,\\
P^{AA}_{\mathrm{ct, \;\ell = 2}}(k) &=& \frac{4}{21}\frac{k^2}{k_{\rm norm}^2}P_{\rm lin}(k)\left[7fc^A_{\mathrm{ct},20} + \left(6f + 7b^A_1\right)c^A_{\mathrm{ct},22}\right.\\
&+&\left.   \left(5f + 6b^A_1\right) \left(c^{A}_{\mathrm{ct},24}+ \frac{k^2}{k_{\rm norm}^2}c^A_{\mathrm{ct},44})\right) + \frac{20}{33}\left(28f + 33b^A_1\right)\left(c^A_{
\mathrm{ct},26
} + \frac{k^2}{k_{\rm norm}^2}c^A_{
\mathrm{ct},46
}\right)\right]\;,\nonumber \\
P^{AA}_{\mathrm{ct, \;\ell = 4}}(k) &=& \frac{16}{385}\frac{k^2}{k_{\rm norm}^2}P_{\rm lin}(k)\left[ 11 f c^A_{\mathrm{ct},22} + \left(15f + 11b^A_1\right)\left(c^A_{\mathrm{ct},24} +\frac{k^2}{k_{\rm norm}^2}c^A_{\mathrm{ct},44}\right)\right.\\
&+&\left. \frac{15}{13}\left(14f + 13b^A_1\right)\left(c^A_{\mathrm{ct},26} + \frac{k^2}{k_{\rm norm}^2}c^A_{\mathrm{ct},46}\right)\right]\;. \nonumber
\end{eqnarray}
We extract each non-linear term of the non-linear spectrum using the publicly available code \texttt{CLASS-PT} \cite{chudaykin2020nonlinear} (including the IR-resummation based on \cite{Senatore:2014via}). Finally, we can already foresee the importance of the $\mu^6$ terms: that there are in total six free terms $c^A_{{\rm ct},20}$, $c^A_{{\rm ct},22}$, $c^A_{{\rm ct},24}$, $c^A_{{\rm ct},26}$, $c^A_{{\rm ct},44}$ and $c^A_{{\rm ct},46}$ to contemplate the two possible scalings ($k^2P_{\rm lin }$ and $k^4P_{\rm lin }$) for the three multipoles considered here.

\subsection{Multi-tracer (MT)}

\paragraph{Real space}

In this section, we generalize the ST expressions to multiple tracers both in real and redshift space. When having two tracers $A$ and $B$, their auto-spectra will be the same as those defined in the previous section. Nevertheless, the cross-power spectrum will have a different form \cite{Mergulhao:2021kip}:
\begin{eqnarray}
     P^{AB}(k) &=& b^A_1b^B_1\left[ P_{\rm lin}(k) + P_{\rm 1L}(k) \right] + \frac{1}{2}\left(b^A_1b^B_2 + b^B_1b^A_2\right)\mathcal{I}_{\delta^2}(k)   \nonumber \\
    &+& \left(b^A_1b^B_{\mathcal{G}_2} + b^B_1b^A_{\mathcal{G}_2}\right)\mathcal{I}_{\mathcal{G}_2}(k)
     + \left[\left(b^A_1b^B_{\mathcal{G}_2} + b^B_1b^A_{\mathcal{G}_2}\right) 
      + \frac{2}{5}\left(b^{A}_1b^B_{\Gamma _{3}} + b^{B}_1b^A_{\Gamma _{3}}\right)\right]\mathcal{F}_{\mathcal{G}_2}(k) \nonumber \\ 
      &+& \frac{1}{4}b^A_2b^B_2\mathcal{I}_{\delta^2\delta^2}(k)  
      + b^B_{\mathcal{G}_2}b^A_{\mathcal{G}_2}\mathcal{I}_{\mathcal{G}_2\mathcal{G}_2}(k) + \frac{1}{2}(b^A_2b^B_{\mathcal{G}_2} + b^B_2b^A_{\mathcal{G}_2} )\mathcal{I}_{\delta^2\mathcal{G}_2}(k)
\nonumber \\ 
&+&P_{\rm ct}^{AB}(k) +  P_{\varepsilon^A\varepsilon^B}(k) \,,
\end{eqnarray}
where we have for $P_{\rm ct}^{AB}$ and for the cross-stochastic term $P_{\varepsilon^A\varepsilon^B}$
\begin{eqnarray}
    P_{\rm ct}^{AB}(k) &=& -k^2P_{\rm lin}(k)
    \left[ 
    2\frac{c_s^2b^A_1b^B_1}{k_{\mathrm{NL}^2}} + b^A_1\left(R^B_{*}\right)^2 + b^B_1\left(R^A_{*}\right)^2 \right]\nonumber  \\
    &=& -\left(b^A_{\nabla^2\delta}b^B_1+b^B_{\nabla^2\delta}b^A_1\right)k^2P_{\rm lin} \; , \\
       P_{\varepsilon^A\varepsilon^B}(k) &=& \frac{1}{\sqrt{\bar{n}_A\bar{n}_B}}\left[c^{AB}_{\mathrm{st},0} + c^{AB}_{\mathrm{st},2} \frac{k^2}{k^2_{\mathrm{norm}}} \right] \,.
\end{eqnarray}
The cross-stochastic term in MT is often neglected since the stochastic contribution depends on the formation history of each tracer on very small scales and is therefore uncorrelated \cite{Desjacques2016}. However, exclusion effects and satellite
galaxies effects \cite{Baldauf:2013hka} may lead to non-Poissonian noise and increase the cross-correlation between the stochastic components for tracers $A$ and $B$ generating e.g. terms proportional to powers of $k^2$. In the first work \cite{Mergulhao:2021kip}, we found a cross-stochastic term consistent with zero and that the inclusion of that term does not deteriorate parameter extraction. In this work, again, we investigate the impact of that cross-stochastic term, which turns out to be very mild (see Sec.~\ref{sec:extra_terms}). 

\paragraph{Redshift space}

After symmetrizing Eq.~(\ref{eq:rsd_st}), we get
\begin{eqnarray}
    P^{AB}(k,\mu) &=& Z^A_1(k)Z^B_1(k)
P_{\rm lin}
(k)+ 2\int_{\q}Z^A_2(\q,\k-\q)Z^B_2(\q,\k-\q)
P_{\rm lin}(z,|\k-\q|)
P_{\rm lin}(q)  \n
&+&  3Z_1^A(\k)P_{\rm lin}(k)\int_{\q}Z_3^B(\q,-\q,\k)P_{\rm lin}(q) +  3Z_1^B(\k)P_{\rm lin}(k)\int_{\q}Z_3^A(\q,-\q,\k)P_{\rm lin}(q) \n
&+& P_{\text{ct}}^{AB}(k,\mu)
 + P_{\varepsilon^A\varepsilon^B}(k,\mu)\,,
\end{eqnarray}
where we have for the counter-terms 
\begin{eqnarray}\label{eq:ct_rsd_mt}
   P_{\rm ct}^{AB}(k,\mu) &=&  \frac{k^2}{k_{\rm norm}^2}P_{\rm lin}(k)\left[Z^A_1\left(c^B_{{\rm ct},20} + c^B_{{\rm ct},22}\mu^2 + c^B_{{\rm ct},24}\mu^4 + c^B_{{\rm ct},44}\frac{k^2}{k_{\rm norm}^2}\mu^4 + \right.\right.\\
   &+& \left.\left. c^B_{{\rm ct},26}\mu^6 + c^B_{{\rm ct},46}\frac{k^2}{k_{\rm norm}^2}\mu^6\right) + A\leftrightarrow B \right]\,, \nonumber
\end{eqnarray}
and for the stochastic term
\begin{eqnarray} \label{eq:cross_stoch_rs}
   P_{\varepsilon^A\varepsilon^B}(k,\mu) &=& \frac{1}{\sqrt{\bar{n}_A\bar{n}}_B}\left[c^{AB}_{\rm st, 00} + c^{AB}_{\rm st, 20}\frac{k^2}{k_{\rm norm}^2} + c^{AB}_{\rm st, 22}\frac{k^2}{k_{\rm norm}^2}f\mu^2 \right]  \,.
\end{eqnarray}
We study the impact of including this cross-stochastic term in Sec.~\ref{sec:extra_terms}.

The counter-terms for the cross-power spectrum, after using the projection~(\ref{eq:project}), are
\begin{eqnarray}
P^{AB}_{\mathrm{ct, \;\ell = 0}}(k) &=& \frac{1}{105}\frac{k^2}{k_{\rm norm}^2}P_{\rm lin}(k)\left[35\left(f + 3 b^B_1\right)c^A_{\mathrm{ct},20} + 7\left( 3f + 5b^B_1\right)c^A_{\mathrm{ct}, 22}  \right.\\ 
 &+& \left. 3\left(5f + 7b^B_1\right)\left(c^A_{\mathrm{ct},24} + \frac{k^2}{k_{\rm norm}^2}c^{A}_{\mathrm{ct}, 44} \right) + \frac{35}{21}\left(7f + b^B_1\right)\left(c^A_{\mathrm{ct},26} + \frac{k^2}{k_{\rm norm}^2}c^A_{\mathrm{ct},46}\right) \right. \nonumber\\
 &+&\left. A \leftrightarrow B\right]\;, \nonumber \\
 P^{AB}_{\mathrm{ct, \;\ell = 2}}(k) &=& \frac{2}{21}\frac{k^2}{k_{\rm norm}^2}P_{\rm lin}(k)\left[7fc^A_{\mathrm{ct},20} + \left(6f + 7b^B_1\right)c^A_{\mathrm{ct},22}  \right.\\
&+&\left. \left(5f + 6b^B_1\right) \left(c^{A}_{\mathrm{ct},24}+ \frac{k^2}{k_{\rm norm}^2}c^A_{\mathrm{ct},44})\right) + \frac{20}{33}\left(28f + 33b^B_1\right)\left(c^A_{
\mathrm{ct},26
} + \frac{k^2}{k_{\rm norm}^2}c^A_{
\mathrm{ct},46
}\right)  \right. \nonumber\\ 
&+&\left. A \leftrightarrow B\right]\;, \nonumber \\
P^{AB}_{\mathrm{ct, \;\ell = 4}}(k) &=& \frac{8}{385}\frac{k^2}{k_{\rm norm}^2}P_{\rm lin}(k)\left[ 11 f c^A_{\mathrm{ct},22} + \left(15f + 11b^B_1\right)\left(c^A_{\mathrm{ct},24} +\frac{k^2}{k_{\rm norm}^2}c^A_{\mathrm{ct},44}\right) \right.  \\
&+&\left. \frac{15}{13}\left(14f + 13b^B_1\right)\left(c^A_{\mathrm{ct},26} + \frac{k^2}{k_{\rm norm}^2}c^A_{\mathrm{ct},46}\right) + A\leftrightarrow B\right]\;. \nonumber
\end{eqnarray}
The full set of parameters in redshift space (in addition to the cosmological parameter) for MT is then:
\begin{eqnarray} \label{eq:all_terms}
   &\left\{ b^A_1,\, b^A_2,\, b^A_{\mathcal{G}_2},\, b^A_{\Gamma_3},\,c^A_{{\rm ct},20},\,c^A_{{\rm ct},22},\,c^A_{{\rm ct},24},\,c^A_{{\rm ct},26},\,c^A_{{\rm ct},44},\,c^A_{{\rm ct},46},\, c^{AA}_{\mathrm{st}, 00},\, c^{AA}_{\mathrm{st}, 02},\, c^{AA}_{\mathrm{st}, 22},\, \right.
    \nonumber \\
    &b^B_1,\, b^B_2,\, b^B_{\mathcal{G}_2},\, b^B_{\Gamma_3},\,c^B_{{\rm ct},20},\,c^B_{{\rm ct},22},\,c^B_{{\rm ct},24},\,c^B_{{\rm ct},26},\,c^B_{{\rm ct},44},\,c^B_{{\rm ct},46},\, c^{BB}_{\mathrm{st}, 00},\, c^{BB}_{\mathrm{st}, 02},\, c^{BB}_{\mathrm{st}, 22}, \nonumber \\  &\left. c^{AB}_{\mathrm{st}, 00},\, c^{AB}_{\mathrm{st}, 02},\, c^{AB}_{\mathrm{st}, 22}\right\}\;.
\end{eqnarray}
The inclusion of that many new free coefficients slows down the MCMC computations. We make use of a Taylor expansion over the cosmological parameters for the computation of the linear power spectrum, similar to Appendix~A of \cite{Mergulhao:2021kip}. We comment more on that in Sec.~\ref{sec:priors}. As we discuss later in the text, the high number of free parameters for MT does {\it not} deteriorate the constraints, since those parameters are more orthogonal and better determined.

\subsection{Combining tracer bias coefficients } \label{sec:combined_tracer}

Since the bias and stochastic parameters encapsulate information from different fields in the cases of MT and ST, it does not make sense to compare them directly. 
When comparing ST and MT, it is useful to define the {\it effective tracer} \cite{Mergulhao:2021kip}.
This {\it effective tracer} is a way to provide a direct (and fair) comparison between the parameters extracted in ST and MT. We define the effective tracer overdensity as
\begin{equation}
    \delta^{\rm eff} = \frac{1}{\bar{n}} 
    \sum_i \bar{n}^i\delta^i \; ,
\end{equation}
where $\bar{n} = \sum_i \bar{n}^i$. Therefore, the effective parameters when considering two tracers are a weighted mean of each tracer coefficient
\begin{eqnarray}
    \label{eqn: effective_bias}
    b^{\mathrm{eff}}_{\left[\boldsymbol{\mathcal{O}}\right]} 
    &=& \frac{1}{\bar{n}} 
    \left( 
    \bar{n}^A b^{\mathrm{A}}_{\left[\boldsymbol{\mathcal{O}}\right]} + 
    \bar{n}^B b^{\mathrm{B}}_{\left[\boldsymbol{\mathcal{O}}\right]} \right) \,,
    \\
    \label{eqn: effective_stoc}
    c^{\mathrm{eff}} &=&
    \frac{1}{\bar{n}} 
    \left[
    \bar{n}^A  c^{AA} 
    + \bar{n}^B c^{BB}+ 2\sqrt{\bar{n}^A\bar{n}^B} c^{AB}
    \right] \, .
\end{eqnarray}
Unless stated otherwise, we refer to the free coefficients of the effective tracer when showing the MT bias and stochastic parameters in Sec.~\ref{sec:results}.


\begin{figure}[h]
    \centering
    \includegraphics[align=c, width = .49\textwidth]{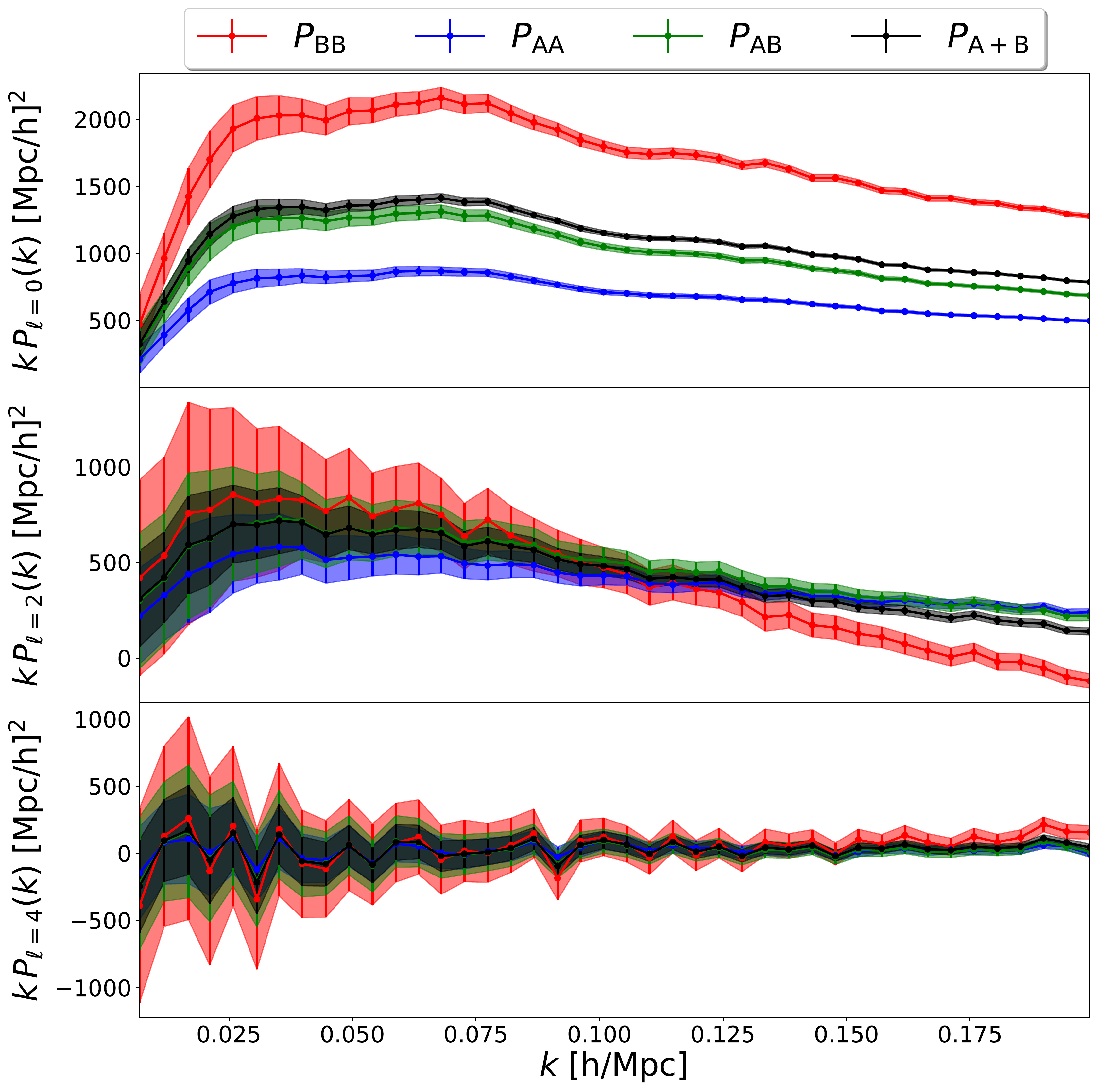}
    \includegraphics[align=c, width = .49\textwidth]{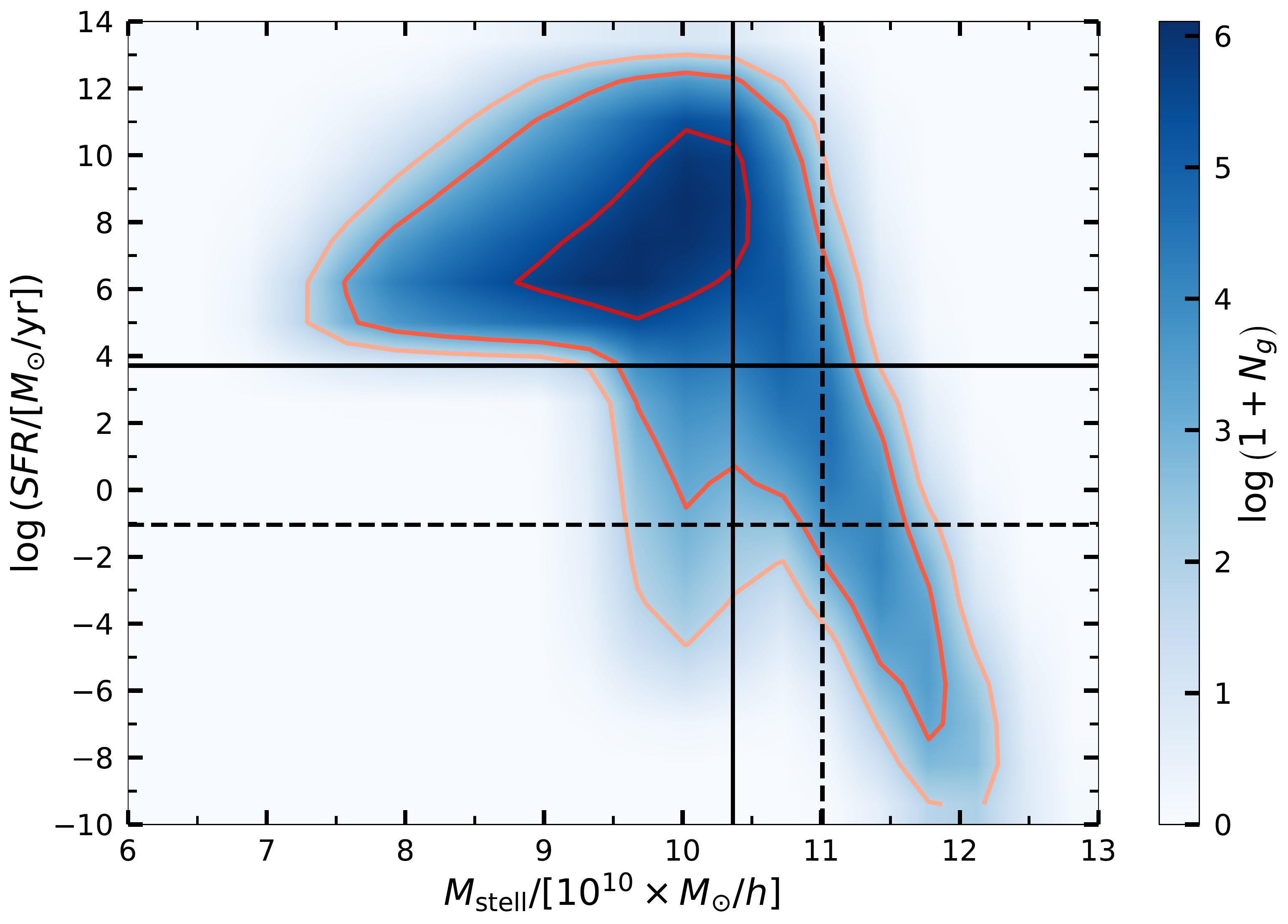}
    \caption{On the left, the power spectrum multipoles ($\ell = 0,2,4$) measured in the simulation for $\bar{n} = \bar{n}_{\rm low}$. We show the auto spectra of tracers $A$ and $B$, their cross-power spectrum $AB$, and the single-tracer case $A+B$. Notice the suppression in the spectrum for the quadrupole of $B$, indicating that the effect of FoG for the tracer $B$ is stronger than for $A$. On the right, the galaxy number $N_g$ as a function of the stellar formation rate (SFR) and the stellar mass, $M_{\rm stell}$. The red lines indicate the level curves $0.3$, $0.5$ and $0.9$ with respect to the peak. The dark solid and dashed lines line shows the cut for the sample with $\bar{n}_{\rm high} = 0.003\, [h/{\rm Mpc}]^3$ and $\bar{n}_{\rm low} = 0.0003\, [h/{\rm Mpc}]^3$, respectively.}
    \label{fig:simulation}
\end{figure}

\section{Data and MCMC setup}
\label{sec:data}

In this section, we start by describing the simulation data. Later, we discuss the covariance matrix in redshift space for multi-tracer, the priors and the MCMC setup. 

\subsection{Simulation} \label{sec:Sim}
We use the dark matter simulations \texttt{TheOne} from the \texttt{BACCO} project \cite{Angulo:2020vky, Zennaro:2021bwy}, in a $1.440\,\mathrm{Gpc}/h$ box with $4320^3$ particles at $z = 0$. The fiducial cosmological parameters  are $\Omega_{\mathrm{cdm}} = 0.259$, $\Omega_{\mathrm{b}} = 0.048$, $h = 0.68$, $n_{s} = 0.96$ and $\sigma_{8} = 0.9$. The simulations were performed with the \textit{paired \& fixed} method of \cite{2016MNRAS.462L...1A}, which strongly suppresses the cosmic variance. That allow us to reescale the volume of the sample by adjusting the covariance \cite{Maion:2022yjo}. Since the simulation volume is $V_s \approx 3\,[\mathrm{Gpc}/h]^3 \approx V_{\rm BOSS}/2$, we rescale the covariance to fit the BOSS volume $V_{\rm BOSS} \sim 6\,[\mathrm{Gpc}/h]^3 $ (see e.g. Table~6 of \cite{Desjacques2016}) and consider $V_{\rm BOSS}$ as the fiducial volume of our work. We discuss the covariance in Sec.~\ref{sec:covariance} and the effect of the sample volume in Sec.~\ref{sec:volumeandshot}.

We populate the (sub)halos from \texttt{TheOne} with galaxies using the {\it subhalo abundance matching extended} (SHAMe) empirical  method of \cite{Contreras:2020him}\footnote{For the SHAMe parameters, we used the values for number density $\bar{n} = 0.001\, [h/{\rm Mpc}]^3$ in Table~1 of \cite{Contreras:2020him}.}. After that, in order to provide two different values for the total number density of the galaxies we perform two different cuts in the galaxy catalogue according to their stellar mass $M_{\rm stell}$. We keep only the most massive objects to obtain: $\bar{n}_{\rm high} = 0.003\, [h/{\rm Mpc}]^3$ and  $\bar{n}_{\rm low} = 0.0003\, [h/{\rm Mpc}]^3$, with the latter being close to the number density of both BOSS and DESI LRG samples (see again Table~6 of \cite{Desjacques2016}). We discuss the impact of the total sample shot noise in Sec.~\ref{sec:volumeandshot}. The SHAMe method also attributes a star formation ratio (SFR) parameter to each galaxy assuming the SFR depends on the dark
matter accretion rate of the host halo, akin to \cite{Moster_2018}. Thus, after performing the first cut in $M_{\rm stell}$ to fix $\bar{n}$, we perform a second cut in the SFR to split the sample into two tracers $A$ and $B$ such that $\bar{n}_A = \bar{n}_B = \bar{n}/2$. This cut resembles a color cut since blue (red) galaxies present a larger (lower) SFR \cite{SFRcolor}. Notice that this is a more realistic scenario to test MT gains when compared to our first work \cite{Mergulhao:2021kip}, which has mostly focused on halos split by their masses. We summarize the cuts in stellar mass and SFR in Table~\ref{table:tracer_definition} and on the right panel of Fig.~\ref{fig:simulation}, where we display the number of galaxies $N_g$ as a function of the SFR and the stellar mass $M_{\rm stell}$. The dark solid line shows the cut for the sample with $\bar{n}_{\rm high} = 0.003\, [h/{\rm Mpc}]^3$. The dark dashed line refers to the cut to obtain the sample with $\bar{n}_{\rm low} = 0.0003\, [h/{\rm Mpc}]^3$. Note that the sample with $\bar{n}_{\rm low}$ corresponds to a cut $M_{\rm stell}\gtrsim 11\times10^{10} \, M_\odot/h$ and that the sample with $\bar{n}_{\rm high}$ corresponds to a cut $M_{\rm stell}\gtrsim 10.3\times10^{10} \, M_\odot/h$. 
 \begin{table}[!htb]
     \begin{minipage}{.5\linewidth}
       \centering
      { \small \begin{tabular}{|c||c|c|} \hline 
                Galaxy set &  SFR [$M_\odot/$yr] & $\bar{n}\,[(\textrm{Mpc}/h)^{-3}]$     \\ \hline \hline
         $A$     & $\gtrsim  10^{-4}$ & $0.0015$ \\ \hline
         $B$     & $\lesssim  10^{-4}$ & $0.0015$  \\ \hline
         $A + B$ &  & $0.0030$  \\
     \hline
     \end{tabular} }
     \end{minipage}%
     \begin{minipage}{.5\linewidth}
       \centering
      { \small \begin{tabular}{|c||c|c|} \hline 
                Galaxy set &  SFR [$M_\odot/$yr] & $\bar{n}\,[(\textrm{Mpc}/h)^{-3}]$     \\ \hline \hline
         $A$     & $\gtrsim  10^{-1}$ &  $0.00015$\\ \hline
         $B$     & $\lesssim  10^{-1}$ &  $0.00015$ \\ \hline
         $A + B$ &  &  $0.00030$  \\
     \hline
     \end{tabular} }
     \end{minipage} 
    
     \caption{Galaxy number density for tracers $A$ and $B$ in the MT scenario and for the total tracer $A+B$ density used for ST. We first perform a cut in the stellar mass $M_{\rm stell}$ attributed by SHAMe to get $\bar{n}_{\rm high}$ and $\bar{n}_{\rm low}$. Later, we perform a cut in SFR to split the tracers into two populations $A$ and $B$. The left tabel refers to $\bar{n}_{\rm low} = 0.0003\, [h/{\rm Mpc}]^3$ and the right tabel to $\bar{n}_{\rm high} = 0.003\, [h/{\rm Mpc}]^3$.}
     \label{table:tracer_definition}
 \end{table}

We display on the left panel of Fig.~\ref{fig:simulation} the monopole, quadrupole, and hexadecapole for each tracer, $A$ and $B$, their cross-spectra ($AB$), and the single-tracer scenario ($A+B$). Note that splitting the ST case into two tracers makes evident that the amplitude of the FoG effect for tracer $B$ is stronger than for tracer $A$. We see that as suppression in the spectrum, especially for higher multipoles. We can relate that to the fact that the tracer split we perform is very similar to the color split of a galaxy sample into blue and red galaxies. It is widely known that red samples reside in more virialized overdensities and therefore present larger peculiar velocities when compared to the blue ones \cite{Madgwick:2003bd, Coil:2007jp, BOSS:2016wmc, Hang:2022zyb}. Since part of the multi-tracer samples may experience stronger FoG suppression, it is crucial to study the impact of including higher-order $\mu^2$ counter-terms to account for that. We discuss this impact in Sec.~\ref{sec:extra_terms}.

\subsection{Covariance}
\label{sec:covariance}

We calculate in this section the covariance matrix for multi-tracer in redshift space. This calculation is based on \cite{DiDioCovariance} and in \cite{Abramo:2022qir}, which calculates the covariance for the angular spectrum.\footnote{We acknowledge and are very thankful to Enea di Dio for providing his derivation for the covariance matrix for multi-tracer in redshift space.}
Defining the power spectrum of two (discrete) fields $\delta_{A}$ and $\delta_{B}$ with its cross shot noise included
\begin{equation}
    \langle \delta_{A}(\boldsymbol{k}) \delta_{B}^\ast(\boldsymbol{k}') \rangle = 
    (2\pi)^3 \delta_{D}(\boldsymbol{k} - \boldsymbol{k}') P^{AB}(\boldsymbol{k})\,,
\end{equation}
we can calculate the covariance by trivially generalizing the real space covariance of \cite{Kaiser:1987qv, Sugiyama:2019ike} to multi-tracer \cite{Mergulhao:2021kip}
{\small \begin{equation}
    {\rm Cov} \left[ P^{AB}(\boldsymbol{k}), P^{CD}(\boldsymbol{k}') \right] =  \frac{(2\pi)^3}{V_s}  \left[ \delta_{\rm D}(\boldsymbol{k} - \boldsymbol{k}')P^{AD}(\boldsymbol{k})  P^{BC}(\boldsymbol{k}) + \delta_{\rm D}(\boldsymbol{k} + \boldsymbol{k}')P^{AC}(\boldsymbol{k})  P^{BD}(\boldsymbol{k})\right]\,,
\end{equation}}
where $V_s$ is the sample volume. Notice that we kept only the Gaussian contribution ($PP$), neglecting the trispectrum part \cite{Sugiyama:2019ike, Wadekar:2019rdu}. Since we are interested in scales $k < 0.2\, h/$Mpc (and more concretely focusing on $k \sim 0.14\, h/$Mpc), the non-Gaussian contribution to the covariance is very small \cite{Sugiyama:2019ike, Wadekar:2019rdu, Blot_2019}.  Using the projection~(\ref{eq:project}), we can now write down the covariance for the projected multipoles as
\bea
{\rm Cov} \left[ P^{ AB}_\ell(\boldsymbol{k}), P^{ CD}_{\ell'}(\boldsymbol{k}') \right] 
&=& \frac{\left( 2 \ell +1 \right) \left( 2 \ell' +1 \right)}{4 V_s} 
\frac{\delta_D \left( k -k' \right) }{2 \pi  k^2}
\int d\Omega_{\hat \bk }d\Omega_{\hat \bk'} \mathcal{L}_\ell \left( \mu_k \right) \mathcal{L}_{\ell'} \left( \mu_{k'} \right) 
 \nonumber \\
&&
\left[
P^{AC} \left( \bk \right) P^{BD} \left( - \bk \right) \delta_D \left( \hat \bk - \hat \bk' \right) 
+ P^{AD} \left( \bk \right) P^{BC} \left( - \bk \right) \delta_D \left( \hat \bk + \hat \bk' \right) 
\right]
\nonumber \\
&=& \frac{\left( 2 \ell +1 \right) \left( 2 \ell' +1 \right)}{4 V_s} 
\frac{\delta_D \left( k -k' \right) }{k^2} \sum_{\ell_1 \ell_2}
\int d\mu \mathcal{L}_\ell \left( \mu \right) \mathcal{L}_{\ell_1} \left( \mu \right) \mathcal{L}_{\ell_2} \left( -\mu \right)
 \nonumber \\
&&
\left[
P_{\ell_1}^{AC} \left(k \right) P^{BD}_{\ell_2} \left(  k \right)  
\mathcal{L}_{\ell'} \left( \mu \right) 
+ P^{AD}_{\ell_1} \left( k \right) P^{BC}_{\ell_2} \left( k \right) \mathcal{L}_{\ell'} \left( -\mu \right) 
\right]\,, 
\eea
where $\Omega_{\hat \bk }$ is the integral over the solid angle formed w.r.t the $\hat \bk$ direction. Again, we only kept the $PP$ contribution. We can simplify the Legendre polynomial products by writing 
\bea
\mathcal{L}_{\ell_1} \left( \mu \right) \mathcal{L}_{\ell_2} \left( \mu \right)  &=& \sum_{\ell_3} \left( 2 \ell_3 +1 \right) \left( \begin{array}{ccc} \ell_1 & \ell_2 & \ell_3 \\ 0 & 0 & 0 \end{array} \right)^2 \mathcal{L}_{\ell_3} \left( \mu \right) \,,
\eea
where the $2\times3$ matrices are the Wigner 3-j symbols. Therefore, we have that
{\small
\bea
\int d\mu \mathcal{L}_\ell \left( \mu \right) \mathcal{L}_{\ell'} \left( \mu \right) \mathcal{L}_{\ell_1} \left( \mu \right) \mathcal{L}_{\ell_2} \left(- \mu \right) 
&=& 2 \left( - 1\right)^{\ell_2} \sum_{\ell_3} \left( 2 \ell_3 +1 \right) 
\left( \begin{array}{ccc} \ell_1 & \ell_2 & \ell_3 \\ 0 & 0 & 0 \end{array} \right)^2 
\left( \begin{array}{ccc} \ell & \ell' & \ell_3 \\ 0 & 0 & 0 \end{array} \right)^2 \,, \\
\int d\mu \mathcal{L}_\ell \left( \mu \right) \mathcal{L}_{\ell'} \left( -\mu \right) \mathcal{L}_{\ell_1} \left( \mu \right) \mathcal{L}_{\ell_2} \left(- \mu \right) &=& 
2 \left( - 1\right)^{\ell'+ \ell_2} \sum_{\ell_3} \left( 2 \ell_3 +1 \right) 
\left( \begin{array}{ccc} \ell_1 & \ell_2 & \ell_3 \\ 0 & 0 & 0 \end{array} \right)^2 
\left( \begin{array}{ccc} \ell & \ell' & \ell_3 \\ 0 & 0 & 0 \end{array} \right)^2 \,,
\eea}
which we can use to write the final covariance for MT in redshift space
\bea \label{eq:finalCov}
{\rm Cov} \left[ P^{AB}_\ell(\boldsymbol{k}), P^{ CD}_{\ell'}(\boldsymbol{k}') \right] &=&
 \frac{\left( 2 \ell +1 \right) \left( 2 \ell' +1 \right)}{2 V_s} 
\frac{ \delta_D \left( k -k' \right) }{k^2}
\nonumber \\
&&
\sum_{\ell_1 \ell_2 \ell_3} \left( -1 \right)^{\ell_2}  \left( 2 \ell_3 +1 \right) 
\left( \begin{array}{ccc} \ell_1 & \ell_2 & \ell_3 \\ 0 & 0 & 0 \end{array} \right)^2 
\left( \begin{array}{ccc} \ell & \ell' & \ell_3 \\ 0 & 0 & 0 \end{array} \right)^2 
\nonumber \\
&&
\qquad
\left[ 
P_{\ell_1}^{AC} \left( k \right) P_{\ell_2}^{BD} \left( k \right) 
+ \left( -1 \right)^{\ell'} P_{\ell_1}^{AD} \left( k \right)   P_{\ell_2}^{BC} \left( k \right) 
\right]\,.
\eea
Notice that, since the simulations described in Sec.~\ref{sec:Sim} were performed with the \textit{paired \& fixed} method \cite{2016MNRAS.462L...1A}, we can (artifitially) reescale the volume $V_s$ of the sample considered in the covariance \cite{Maion:2022yjo}.

\subsection{Priors and MCMC setup}
\label{sec:priors}
Table \ref{tab:priors} describes the priors chosen for the parameters considered in our MCMC analysis. The boundaries of the uniform ({\it flat}) priors are based on previous work and preliminary analysis of our pipeline with the data. We checked that all parameter posteriors are within the priors considered (see also \cite{Mergulhao:2021kip} for a very broad discussion on the different choices of priors in the context of MT). On top of the free terms from Eq.~(\ref{eq:all_terms}), we explore the posteriors for three cosmological parameters: the density of cold dark matter $\omega_{\rm cdm} = h^2\Omega_{\rm cdm}$, the Hubble parameter $h$ and the (rescaled) amplitude of fluctuations $A_s \times 10^{9}$.

\begin{table}[!htb]
    \begin{minipage}{.5\linewidth}
      \centering
    \begin{tabular}{|c||c|} \hline 
     Parameter &  \textit{Flat}  Prior  \\ \hline \hline
    $\omega_{\rm cdm}$  &  $[0.09, 0.14]$\\ \hline
    $h$   & $[0.57, 0.77]$ \\ \hline
    $A_s \times 10^{9}$   & $[1.5, 4]$\\ \hline
    \hline
    $b_1$   & $[0.6, 2]$    \\ \hline
    $b_2$   &  $[-10, 10]$ \\ \hline 
    $b_{\mathcal{G}_2}$   &  $[-10, 10]$  \\ \hline 
    $b_{\Gamma_3}$   &  $[-15, 15]$  \\ \hline 
    \end{tabular}
    \end{minipage}%
    \begin{minipage}{.5\linewidth}
      \centering
    \begin{tabular}{|c||c|} \hline 
     Parameter &  \textit{Flat}  Prior  \\ \hline \hline
    $c_{\rm ct,20}$   &  $[-10, 10]$  \\ \hline 
    $c_{\rm ct,22}$   &  $[-10, 10]$  \\ \hline 
    $c_{\rm ct,24}$   &  $[-10, 10]$  \\ \hline 
    $c_{\rm ct,26}$   &  $[-50, 50]$  \\ \hline 
    $c_{\rm ct,44}$   &  $[-10, 10]$  \\ \hline 
    $c_{\rm ct,46}$   &  $[-50, 50]$  \\ \hline 
    \hline
    $c_{\rm st,00}$   &  $[-2, 2]$  \\ \hline 
    $c_{\rm st,02}$   &  $[-20, 20]$  \\ \hline 
    $c_{\rm st,22}$   &  $[-50, 50]$  \\ \hline 
    \end{tabular}
    \end{minipage} 
    
    \caption{Priors on the different parameters considered in our MCMC analysis. For a broader discussion of priors in the context of MT, see \cite{Mergulhao:2021kip}.}
    \label{tab:priors}
\end{table}

We sampled the posterior using the MCMC method and two different \texttt{Python} packages: \texttt{emcee} \cite{Foreman-Mackey:2012any} and \texttt{pocoMC} \cite{karamanis2022pocomc, karamanis2022accelerating}. Sampling the MT likelihood can be challenging, since it has approximately $30$ parameters and some degeneracies. We, therefore, decided to take a conservative approach and cross-checked that these two samplers lead to consistent results. For the \texttt{emcee}, we used $20$ walkers per parameter and used the Gelman-Rubin criteria \cite{Gelman:1992zz} to determine its convergence (with four parallel chains and scale reduction factor $\epsilon < 0.03$). For \texttt{pocoMC}, we used its standard configurations and built-in convergence criteria, with $5000$ particles. Both \texttt{emcee} and \texttt{pocoMC} led to similar posteriors, with some minor differences in some of the nuisance parameters. We used the \texttt{emcee} chains to make the figures in this work.

Finally, each MCMC iteration requires running a numerical Boltzmann solver to get the linear matter power spectrum, and additional loop integrals from perturbation theory. For that, we used \texttt{CLASS-PT} \cite{chudaykin2020nonlinear}, a modified version of \texttt{CLASS} \cite{Blas:2011rf}. Considering the large number of free parameters considered, each iteration of \texttt{CLASS-PT} would be numerically expensive. Therefore, we Taylor expanded all cosmology-dependent terms in our analysis following Appendix~A of \cite{Mergulhao:2021kip}. The Taylor expansion boosts the evaluation of each MCMC iteration by a factor of  $\approx 100$ relative to running \texttt{CLASS-PT}.

\section{Results}
\label{sec:results}
In this section we explain the main results of the paper. We start by discussing the relevance of the higher-in-$\mu$ coefficients in the counter-terms to parametrize the FoG for MT and the importance of the cross-stochastic coefficients. Later, we consider the behaviour of MT for different shot noises and box volumes. In the last part, we explain and quantify the benefits of MT analysis compared to ST. We show results for $z=0$, different box sizes and sample shot noise.

\subsection{Higher-in-$\mu$ and cross-stochastic terms}
\label{sec:extra_terms}

\begin{figure}
    \centering
    \includegraphics[width = 0.49\textwidth]{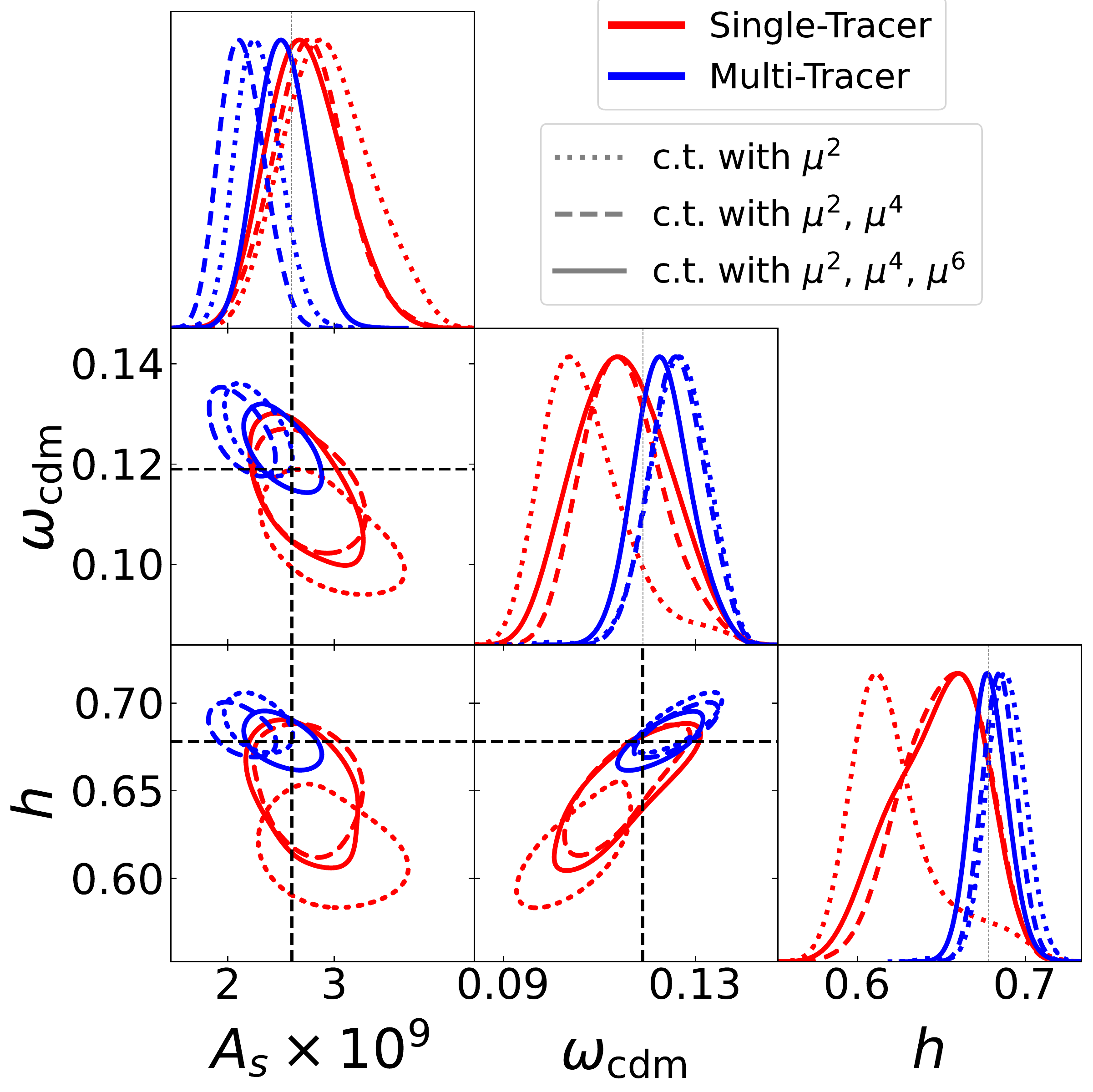}
    \includegraphics[width = 0.49\textwidth]{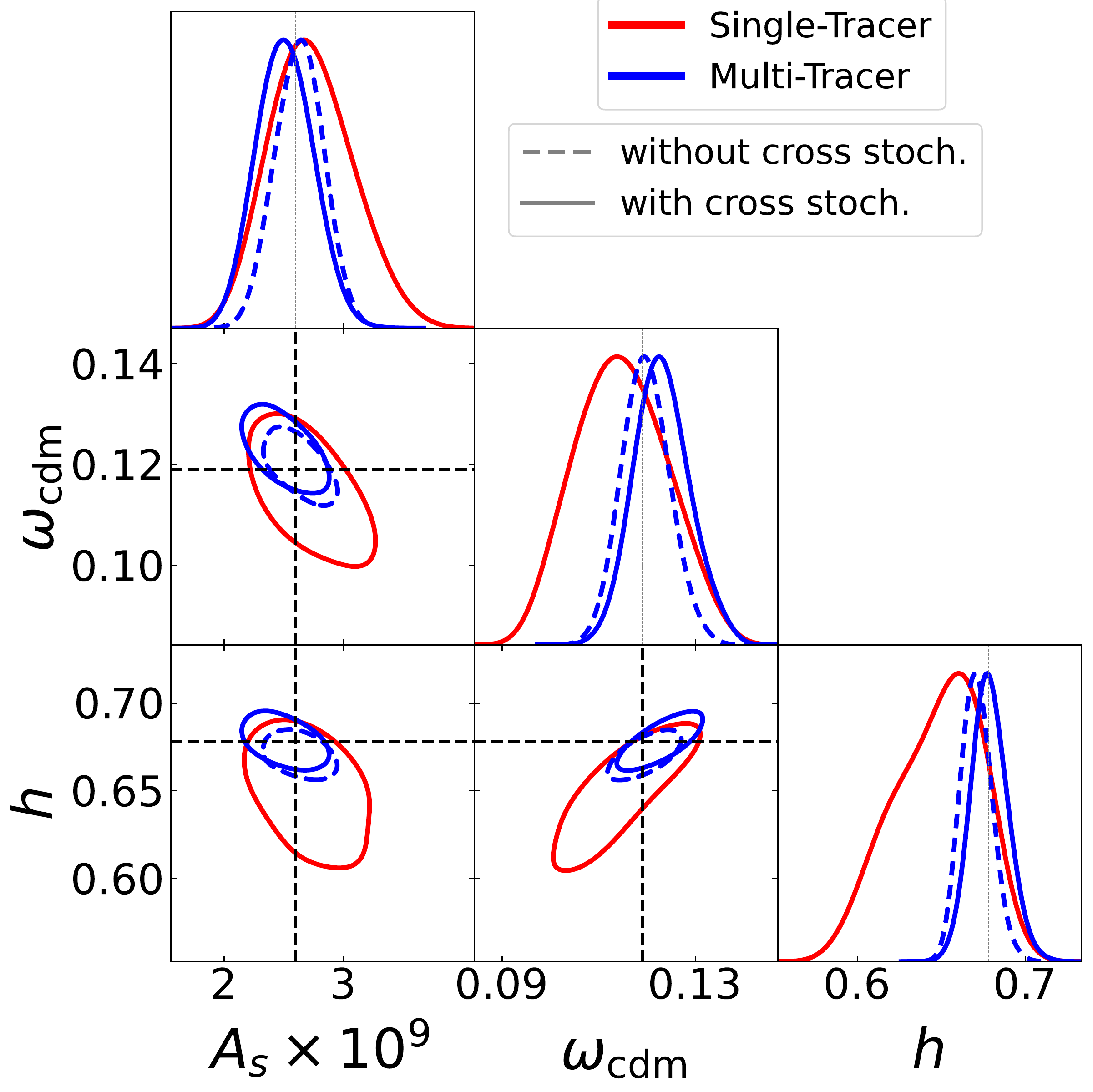}
    \caption{On the left, we show the effect on the parameter posteriors ($1\sigma$) when including higher-in-$\mu$ terms in Eq.~(\ref{eq:ct_rsd}). On the right, the effect of including the cross-stochastic terms of Eq.~(\ref{eq:cross_stoch_rs}) for MT. In both panels we consider $k_{\rm max} = 0.14h/$Mpc and the largest value for the tracer number density $\bar{n}_{\rm high}$.}
    \label{fig:mu6_dedendence}
\end{figure}
We begin by investigating which terms are relevant when considering MT. 
On the left panel of Fig.~\ref{fig:mu6_dedendence}, we study the relevance of higher-order-in-$\mu$ coefficients for the counter-term in Eqs.~(\ref{eq:ct_rsd}) and (\ref{eq:ct_rsd_mt}) when gradually adding the higher-in-$\mu$ parameters. We fix the maximum Fourier mode that was used in the analysis to $k_{\rm max} = 0.14h/$Mpc and the total tracer number density $\bar{n} = \bar{n}_{\rm high}$. Moreover, all the other terms in Eq.~(\ref{eq:all_terms_ST}) for ST and (\ref{eq:all_terms}) for MT are included. Notice that for ST (red lines), including $\mu^4$ is vital, shifting the posterior towards the right direction. It agrees with the analysis of other works \cite{chudaykin2020nonlinear, Ivanov:2019pdj, Nishimichi:2020tvu} that found those terms to be relevant as a proxy for FoG effects. For MT (blue lines), we notice that the inclusion of a term proportional to $\mu^6$ is also important. Neglecting this term for MT leads to a small shift ($1.5\sigma$) in the amplitude of fluctuations $A_s$. We checked that this bias persists for different values of $k_{\rm max}$. The need for those extra counter-terms is to some extend expected, since the split into different tracers can uncover non-linear behaviour that were smoothed out from the power spectrum of the $A+B$ sample. We can see that, for instance, different intensities of FoG suppression for tracers $A$ and $B$ (see left panel of Fig.~\ref{fig:simulation}). Including high-order-in-$\mu$ terms is therefore vital to describe the tracer $B$ population, which presents stronger FoG suppression. We highlight again that the importance of the $\mu^6$ terms is to contemplate the two possible scalings ($k^2P_{\rm lin }$ and $k^4P_{\rm lin }$) for the three multipoles considered (total of six degrees of freedom per tracer) with six free coefficients per tracer (see Eq.~(\ref{eq:ct_rsd})). Notice, however, that the inclusion of the $\mu^6$ term does not deteriorate either ST or MT posteriors, which motivates us to keep this term in further analysis. We comment on the degeneracy between those terms later in the paper.

Another relevant set of contributions to be considered for MT is the stochastic term for the cross-spectra of MT. We study on the right panel of Fig.~\ref{fig:mu6_dedendence} the effect of including the stochastic terms, following Eq.~(\ref{eq:cross_stoch_rs}). As explained before, this term is often neglected since the tracers $A$ and $B$ do not correlate on small scales. However, exclusion effects may lead to a non-vanishing cross-stochastic contribution \cite{Baldauf:2013hka}. We see that the inclusion of this term does {\it not} lead to any substantial difference when constraining the cosmological parameters, confirming the results in real space \cite{Mergulhao:2021kip}. The posteriors when including this cross-stochastic terms (blue solid) are very similar to those when neglecting it (blue dotted), both being very smaller than the ST scenario. For completeness, we include this cross-stochastic contribution for MT in further analysis. Notice that we fixed $k_{\rm max} = 0.14h/$Mpc and $\bar{n}_{\rm high} = 0.003\, [h/{\rm Mpc}]^3$ for Fig.~\ref{fig:mu6_dedendence} but we checked that the result is very similar for the sample with large shot noise and $\bar{n}_{\rm low} = 0.0003\, [h/{\rm Mpc}]^3$. 

\subsection{Survey volume and sample shot noise}
\label{sec:volumeandshot}

\begin{figure}
    \centering
    \includegraphics[width = 0.49 \textwidth]{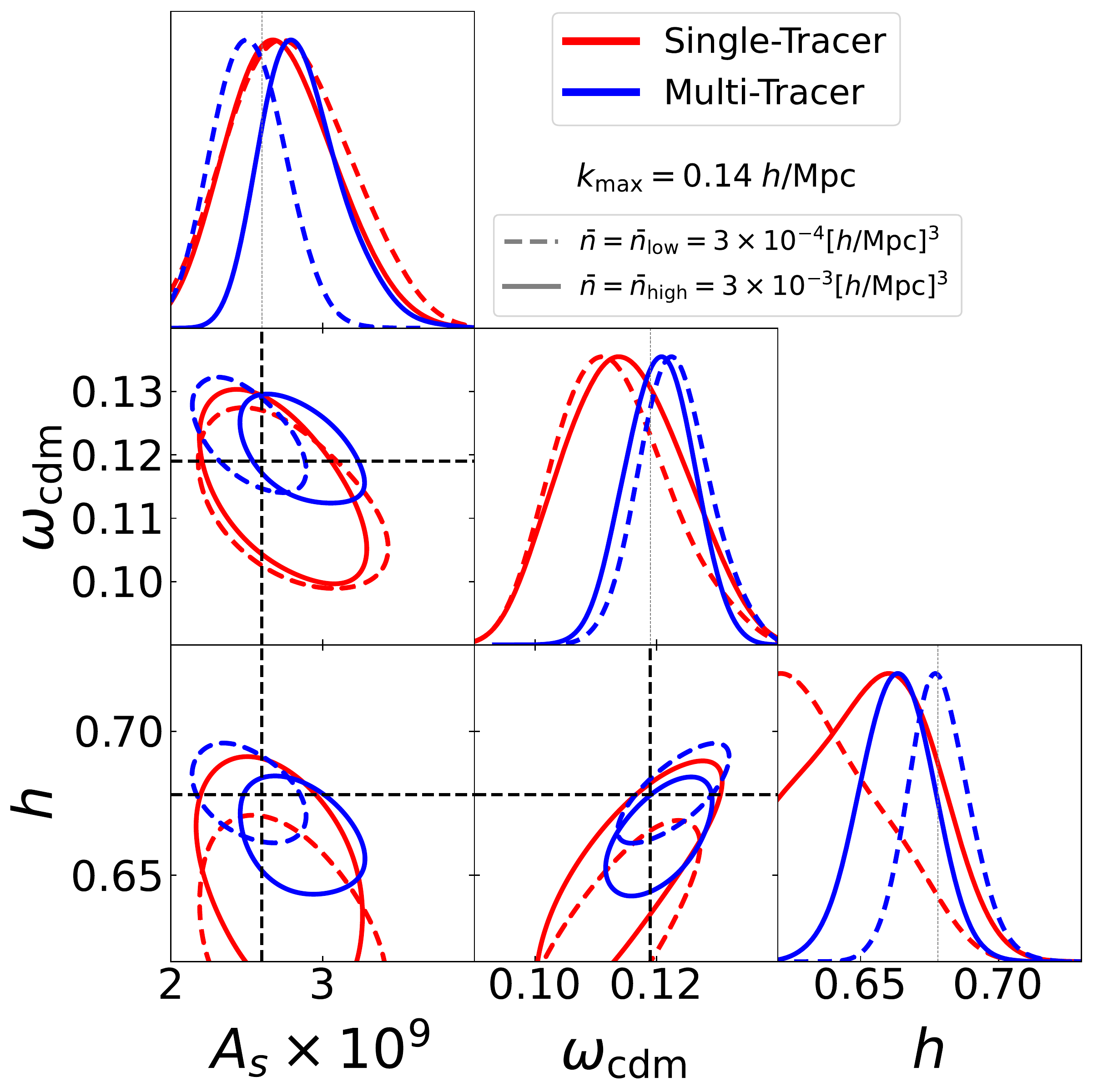}
    \includegraphics[width = 0.49 \textwidth]{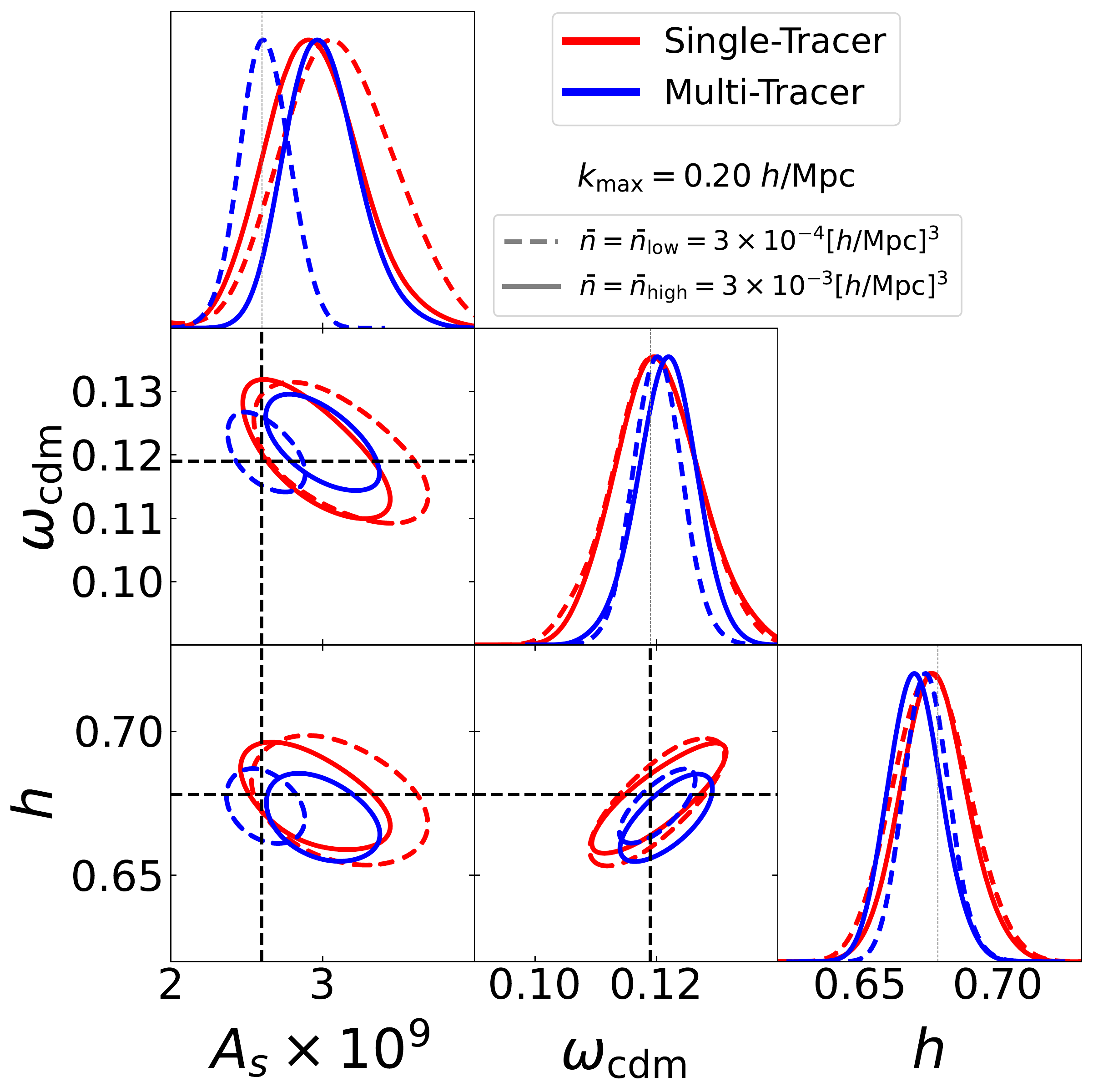}
    \caption{Impact of shot noise on the $1\sigma$ regions of the posterior for $k_{\rm max} = 0.14h/$Mpc (left) and $k_{\rm max} = 0.2h/$Mpc (right). We consider two different values for the tracer number density: $\bar{n} = \bar{n}_{\rm low} = 0.0003\, [h/{\rm Mpc}]^3$ (dashed) that has a larger shot noise and $\bar{n} = \bar{n}_{\rm high} = 0.003\, [h/{\rm Mpc}]^3$ (solid) that has a lower shot noise. }
    \label{fig:shotnoise}
\end{figure}
A potential disadvantage of doing MT is that each sub-sample will have a larger shot noise. In order to check if the shot noise is indeed a limitation, we now show in Fig.~\ref{fig:shotnoise} the impact of the shot noise for multi-tracer and single-tracer on the cosmological parameter posteriors. We use two different values for the total tracer number density, as described in Sec.~\ref{sec:Sim}: $\bar{n}_{\rm high} = 0.003\, [h/{\rm Mpc}]^3$ (solid) and  $\bar{n}_{\rm low} = 0.0003\, [h/{\rm Mpc}]^3$ (dashed). For tracers $A$ and $B$, we have $\bar{n}_{A} = \bar{n}_{B} = \bar{n}/2$. The left panel shows the result for $k_{\rm max} = 0.14h/$Mpc. We notice that the parameter posteriors for MT for different $\bar{n}$ are slightly shifted from each other, but agree on $1\sigma$ with the fiducial values. For ST, the Hubble parameter $h$ presents a small bias ($\sim1\sigma$) for the sample with larger shot noise. On the right panel, we set $k_{\rm max} = 0.2h/$Mpc. In this case, the sample with larger shot noise ($\bar{n}_{\rm low} = 0.0003\, [h/{\rm Mpc}]^3$) presents a $\sim1\sigma$ bias in $A_s$ both for MT and ST. It is important to highlight that, at $z = 0$, higher-order terms from perturbation theory might start to be relevant on those scales. Overall, we can conclude that MT results are consistently better relative to ST for both $\bar{n}_{\rm high}$ and $\bar{n}_{\rm low}$, the latter being close to the shot noise considered for the red galaxies for eBOSS and DESI. The persistence of MT improvements for both number densities indicates that their origin is not solely due to cosmic variance cancellation, since that cancellation is highly dependent on $\bar{n}$ and expected to be more relevant for tracers with large densities \cite{Karagiannis:2023lsj,Seljak:2008xr, Abramo2013}.

Finally, we discuss the perspectives of MT analysis when increasing the survey volume. As explained in Sec.~\ref{sec:data}, the simulation was performed using the \textit{paired \& fixed} method of \cite{2016MNRAS.462L...1A}, which substantially suppresses the cosmic variance. It allow us to safely rescale the sample volume, $V_s \approx 3\,[\mathrm{Gpc}/h]^3$, to higher values by simply adjusting the covariance~(\ref{eq:finalCov}) \cite{Maion:2022yjo}. On the right panel of Fig.~\ref{fig:st_compared_single} we show the constraints from MT and ST for different volumes, which are in units of the volume of the BOSS LRG galaxy sample. For comparison, notice that $V_{s} \approx V_{\rm BOSS}/2 \approx V_{\rm DESI}/13$. The right panel of Fig.~\ref{fig:st_compared_single} indicates that, as expected, increasing the simulation volume led to a substantial reduction in the error bars, without introducing any bias and significant shift in the posterior. In summary, MT method can also provide better constraint for larger sample volumes, such as those from DESI.

\subsection{MT compared to ST}
\label{sec:mt_improvement}

\begin{figure}[h]
    \centering
    \includegraphics[width = 0.49\textwidth]{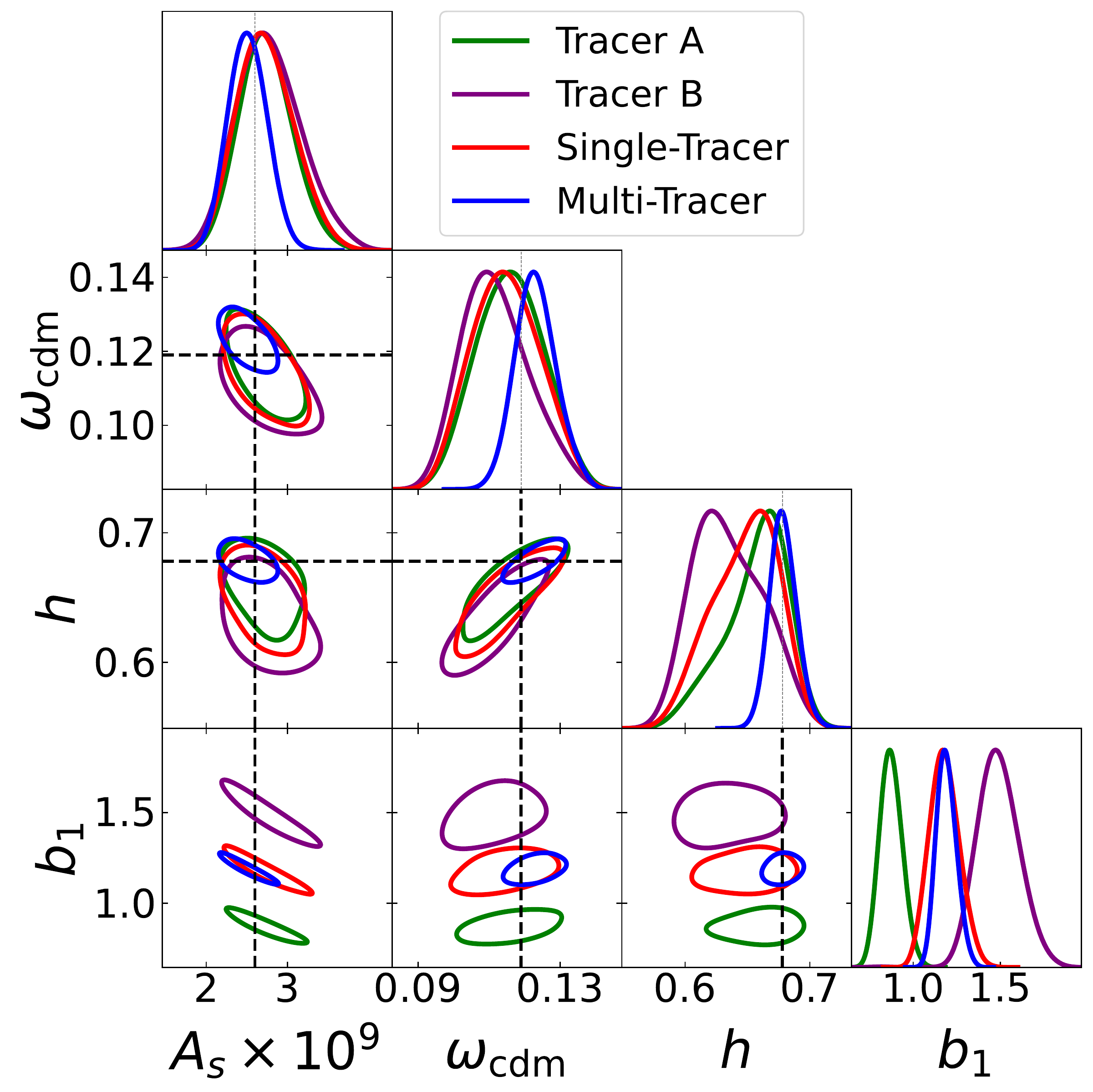}
     \includegraphics[width = 0.49\textwidth]{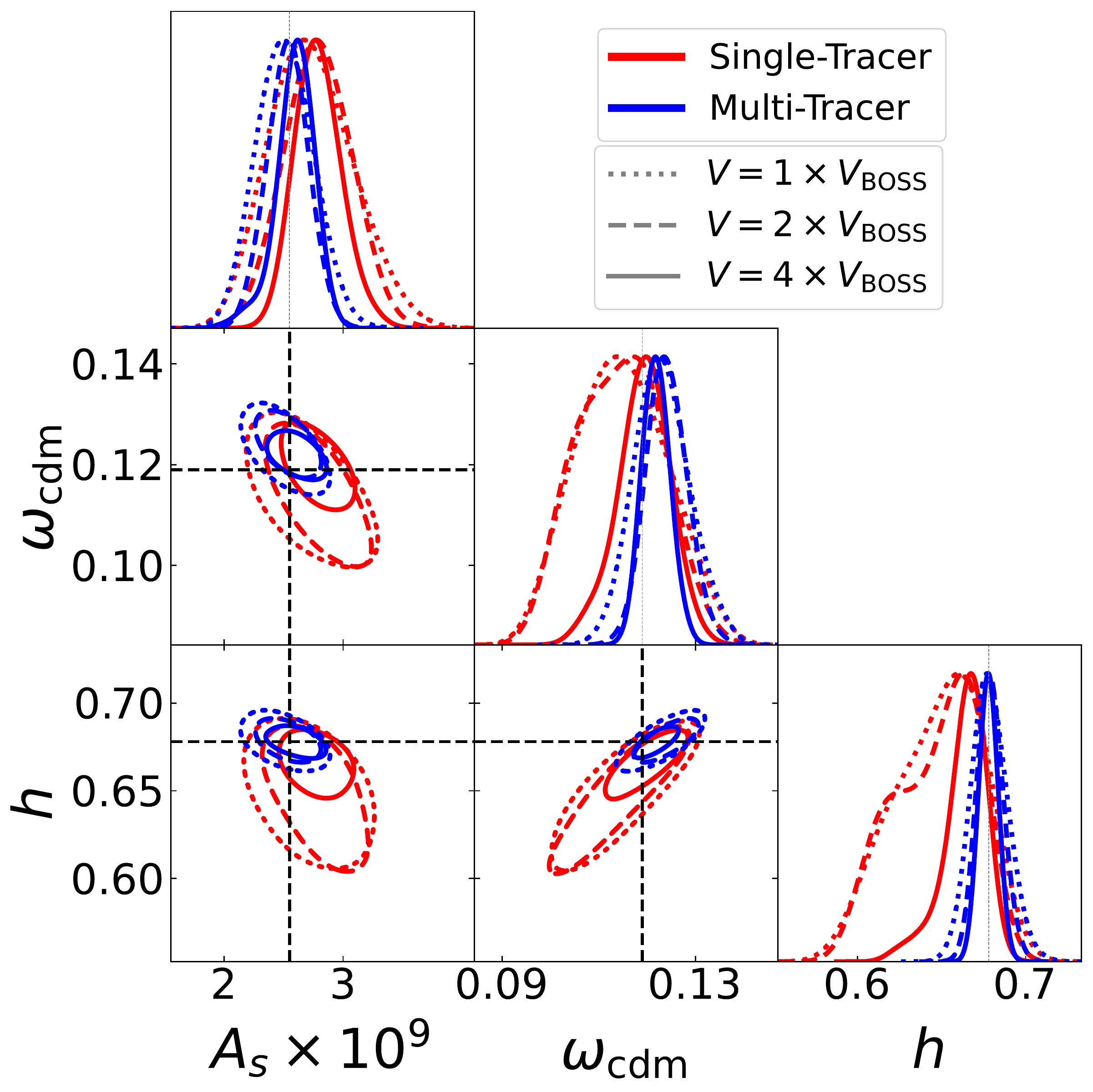}
         \caption{On the left, the cosmological parameter posteriors extracted (at $z=0$) when considering the tracers $A$ and $B$ alone, compared to the MT and ST scenarios. On the right, the cosmological parameters extracted for ST and MT when varying the volume considered for the covariance~(\ref{eq:finalCov}). We fix for those figures $k_{\rm max} = 0.14\,h\mathrm{Mpc}^{-1}$ and $\bar{n} = \bar{n}_{\rm high}$.} 
    \label{fig:st_compared_single}
\end{figure}

We now move to estimate and understand the improvements of MT compared to ST.
We start by comparing on the left panel of Fig.~\ref{fig:st_compared_single}: ST, MT and the constraints extracted when considering the tracers $A$ and $B$ alone. We first notice that MT substantially outperforms the other constraints, while ST (which includes both tracers $A$ and $B$ together) presents comparable constraints as the tracers $A$ and $B$ alone. One would expect that combining the tracers $A$ and $B$ into a single species (ST) would improve the constraints, since the number of objects is increased by a factor of two. However, this is not the case since the shot noise on the scales considered ($k_{\rm max} = 0.14h/$Mpc) is too small compared to the signal (for both $\bar{n}_{\rm low}$ and $\bar{n}_{\rm high}$). It means that the error bars are dominated by the number of modes available in that volume, which is the same for ST, MT and tracers $A$ and $B$ alone. We can therefore conclude that the total sample ($A+B$) presents similar amount of information as the individual sub-samples alone. On the other hand, the fact that MT analysis of these same samples (tracers $A$ and $B$) led to tighter constraints on the cosmological parameter indicates that there is an {\it information gain}. A key source of information that is not present in the ST case is the cross power spectrum between the tracers.
In \cite{Mergulhao:2021kip}, we have shown that the cross power spectrum helps to break degeneracies between bias parameters (e.g., by constraining the product $b_\mathcal{O}^Ab_\mathcal{O}^B$). Throughout this section, we aim to provide more arguments to show that MT can be a way to extract more powerful constraints in redshift space. 

\begin{figure}[h]
    \centering
    \includegraphics[width = 0.49\textwidth]{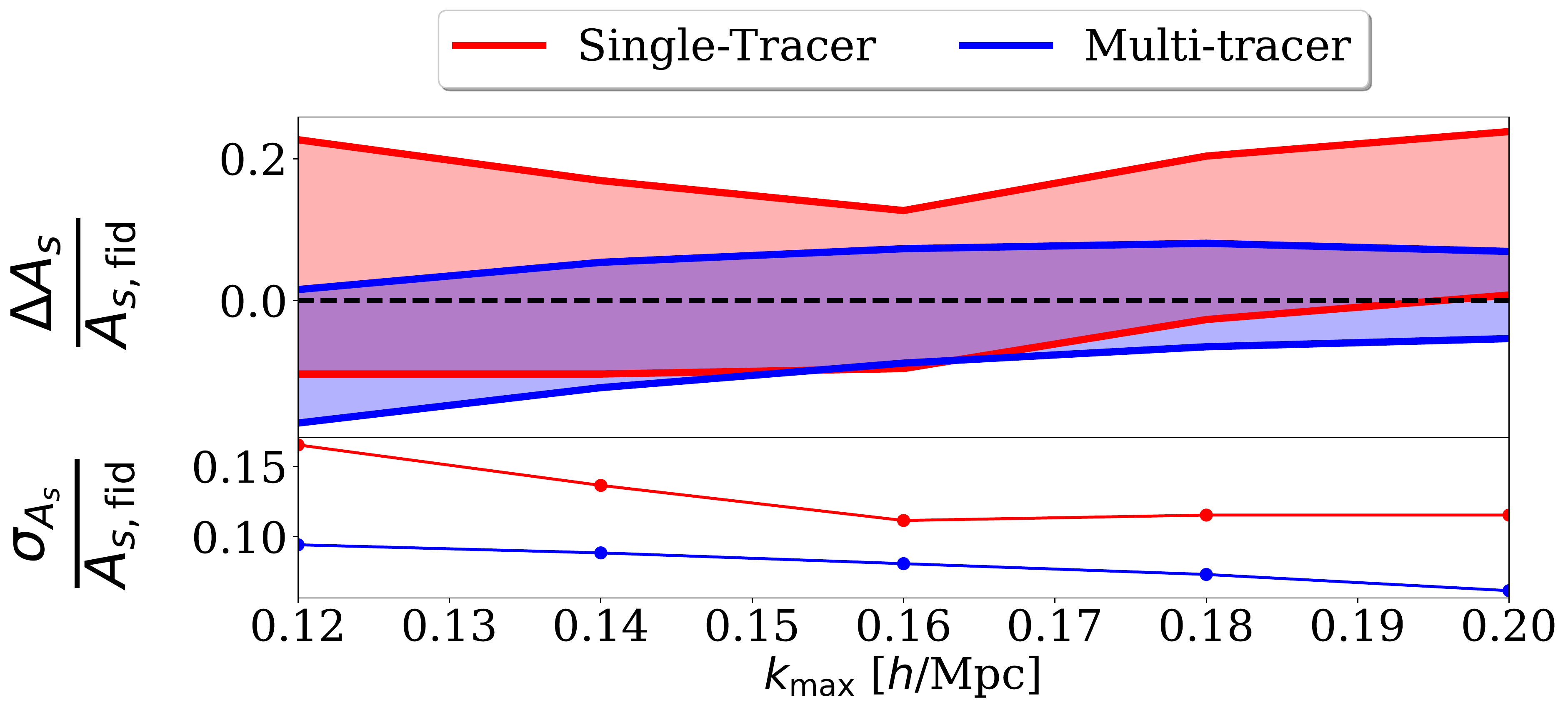}
    \includegraphics[width = 0.49\textwidth]{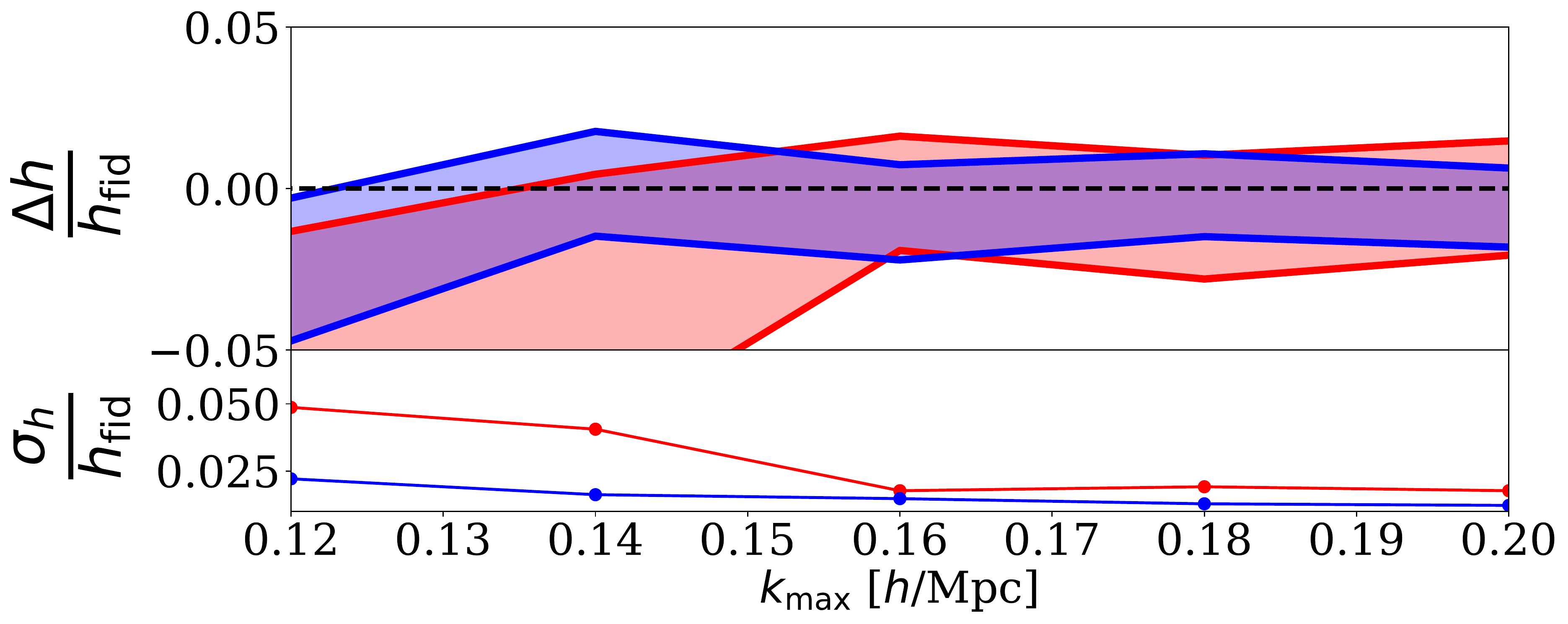}
    \includegraphics[width = 0.49\textwidth]{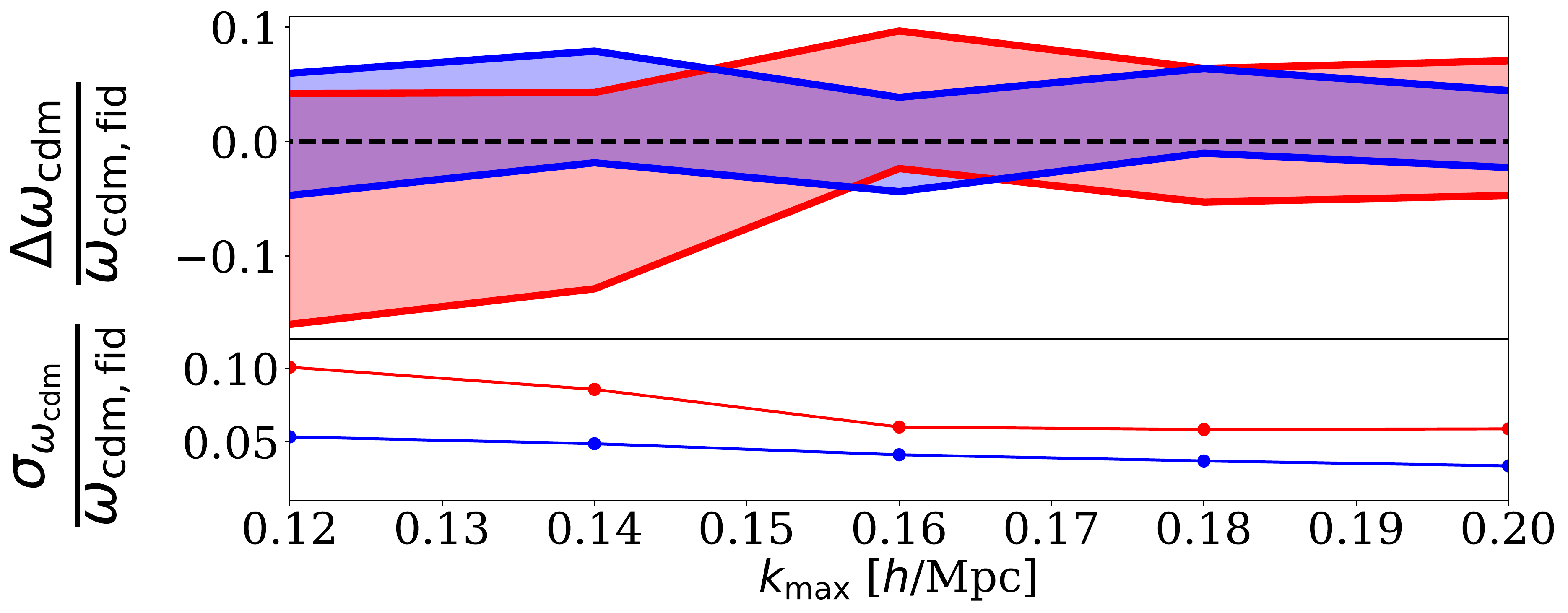}
    \includegraphics[width = 0.49\textwidth]{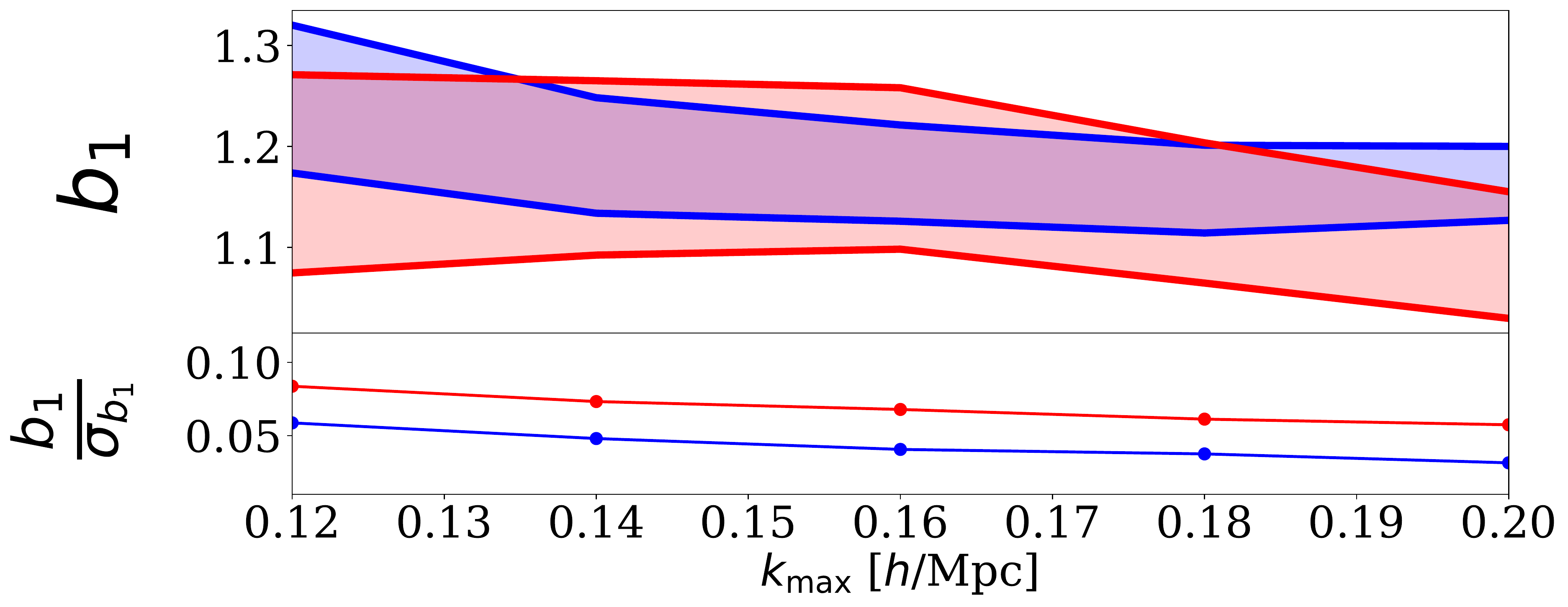}
    \caption{Evolution of the error in different parameters as a function of $k_{\rm max}$ for MT and ST. The shaded regions indicate the error for $A_s$, $\omega_{\rm cdm}$, $h$ and $b_1$. The lines indicate the relative size of the error bars and the values of the cosmological parameters are normalized by the fiducial values from the simulation.}
    \label{fig:kmax_scale}
\end{figure}
To further compare the MT improvements relative to ST, we show in Fig.~\ref{fig:kmax_scale} their results as a function of $k_{\rm max}$, fixing $\bar{n} = \bar{n}_{\rm high}$. For $k_{\rm max} \geq 0.16\,h$/Mpc, we see that the error bars of MT are substantially better: $\sigma_{\rm MT}/\sigma_{\rm ST} \sim 2/3$ for $\omega_{\rm cdm}$, $A_s$ and $h$. For $k_{\rm max} \leq 0.14\,h$/Mpc, the MT improvement is even better, with $\sigma_{\rm MT}/\sigma_{\rm ST} \sim 6/10$ for $\omega_{\rm cdm}$ and $A_s$ and $\sigma_{\rm MT}/\sigma_{\rm ST} < 1/2$ for $h$. At low $k$, the signal is less affected by the increasing shot noise from MT. When increasing $k$, there is a non-trivial $k$ scaling due to how both MT and ST incorporate information and are able to determine the extra free parameters from the EFTofLSS and the bias expansion. After have included the $\mu^4$ terms for ST and $\mu^6$ terms for MT, we do not see any indication of strong bias in the results for the range of $k_{\rm max}$ considered.

\begin{figure}[h]
    \centering
    \includegraphics[width = 0.97\textwidth]{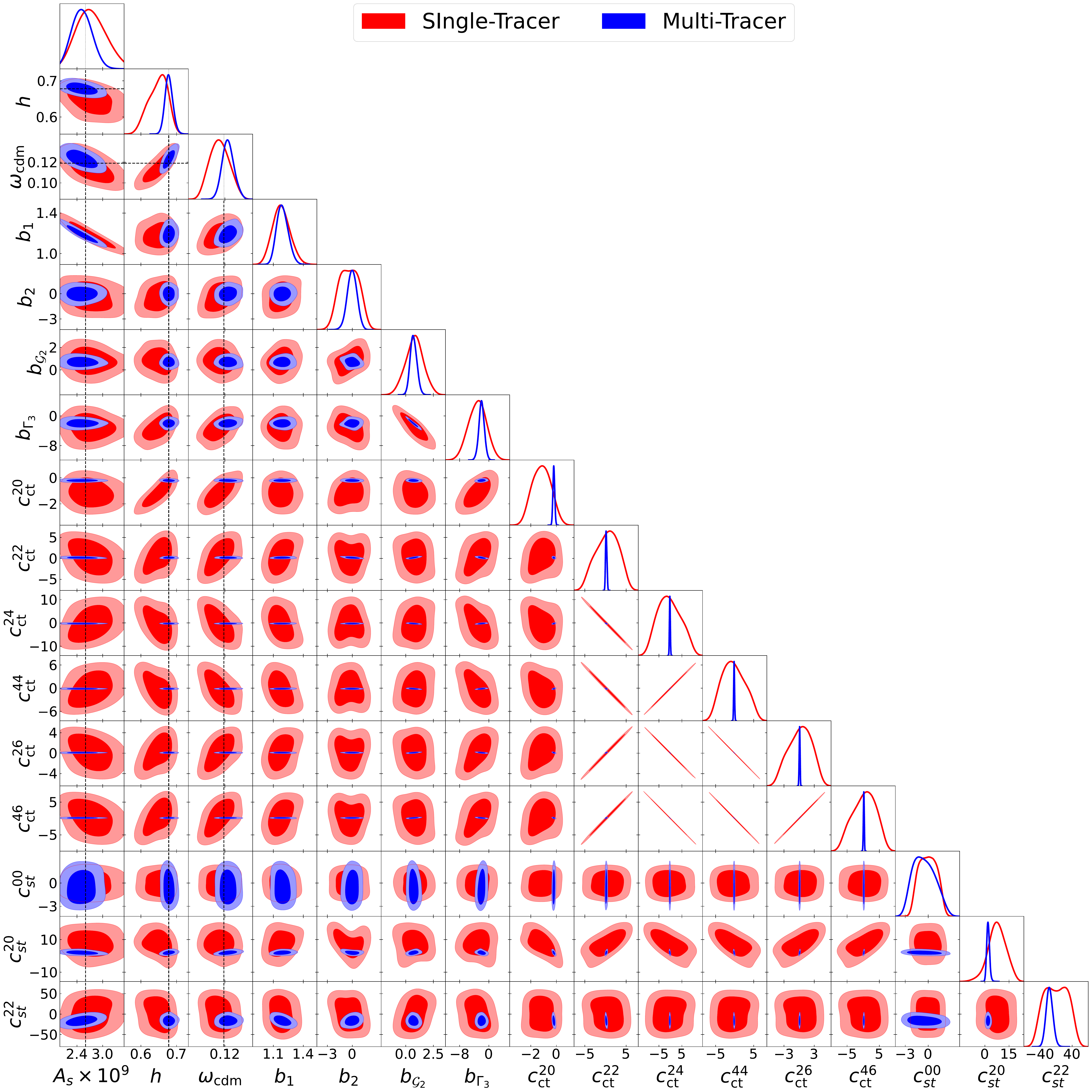}
    \caption{Comparison between MT and ST in redshift space using the full set of parameters in Eqs.~(\ref{eq:ct_rsd}) and (\ref{eq:ct_rsd_mt}) for $k_{\rm max} = 0.14 h/$Mpc, $\bar{n}_{\rm high}$ and $z=0$.} 
    \label{fig:fullmcmc}
\end{figure}

To help to understand why MT outperforms ST, we display in Fig.~\ref{fig:fullmcmc} the MCMC posteriors for all parameters, comparing MT and ST for $k_{\rm max} = 0.14 h/$Mpc and $\bar{n} = \bar{n}_{\rm high}$. As described in Sec.~\ref{sec:combined_tracer}, the MT counter-terms, stochastic and bias parameters displayed refer to the {\it effective tracer} values. Notice that MT outperforms ST for almost all parameters. In addition to the relative gain in the cosmological parameters, we notice a substantial improvement in the bias terms $b_1, b_2, b_{\mathcal{G}_2}$ and $b_{\Gamma_3}$, ranging from a factor 1.5 to 3 better if compared to ST. The stochastic parameters are also better determined for MT. It brings us to the first important conclusion: {\it splitting one population into two tracers has led to a better estimate of the non-linear part of the matter spectrum for each tracer despite having more free parameters}. 

Moreover, Fig.~\ref{fig:fullmcmc} points to a (almost exact) degeneracy between $c_{\rm ct}$'s for ST indicating that a few of those terms are not really necessary in the ST analysis. We include those terms for completeness but, as commented above, we explicitly checked that the results are very similar when dropping those terms from ST. For multi-tracer, on the other hand, they are required in order to have unbiased results for the cosmological parameters (see Fig.~\ref{fig:mu6_dedendence}), despite still being partially degenerate with other terms, as we discuss below. We can also spot a degeneracy between $c_{\rm ct}^{20}$ both with $h$ and $\omega_{\rm cdm}$ for ST, which can partially explain the $1\sigma$ deviation of $h$ seen in Fig.~\ref{fig:kmax_scale} at low $k_{\rm max}$. That may be induced by volume effect since it goes away for larger volumes (see Fig.~\ref{fig:st_compared_single}).  

\begin{figure}[h]
    \centering
    \includegraphics[width = 0.49\textwidth]{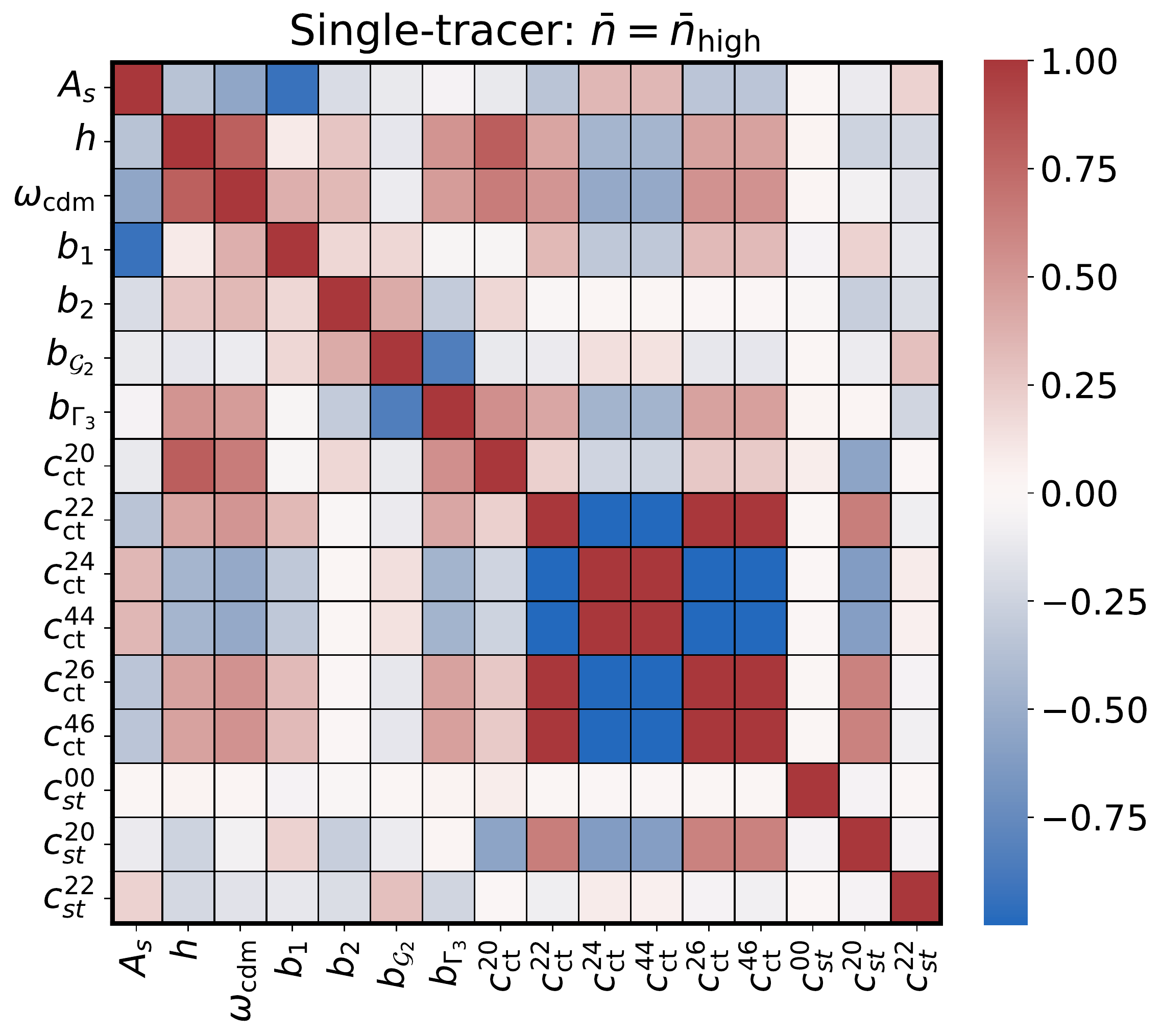}
    \includegraphics[width = 0.49\textwidth]{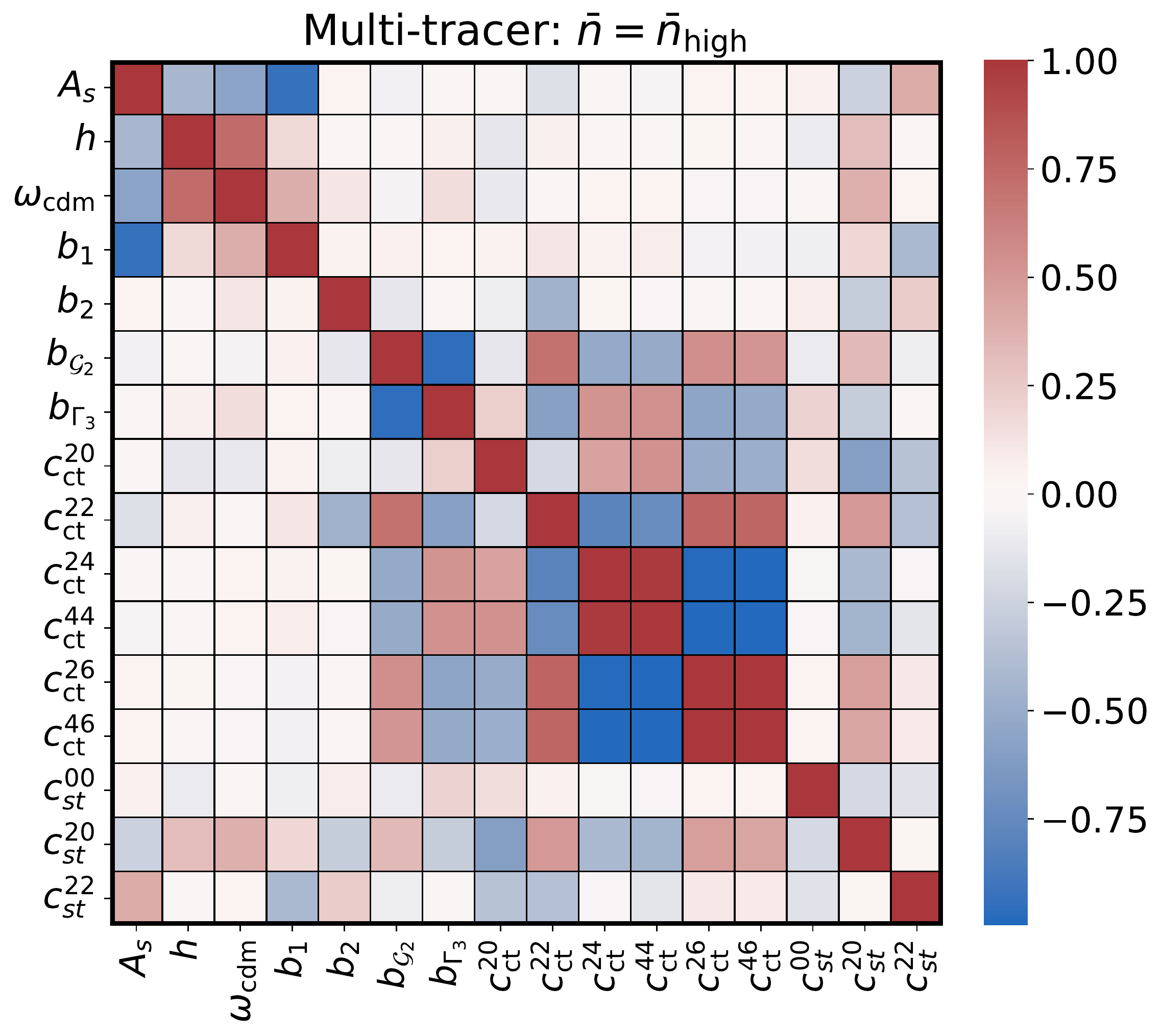}
    \includegraphics[width = 0.49\textwidth]{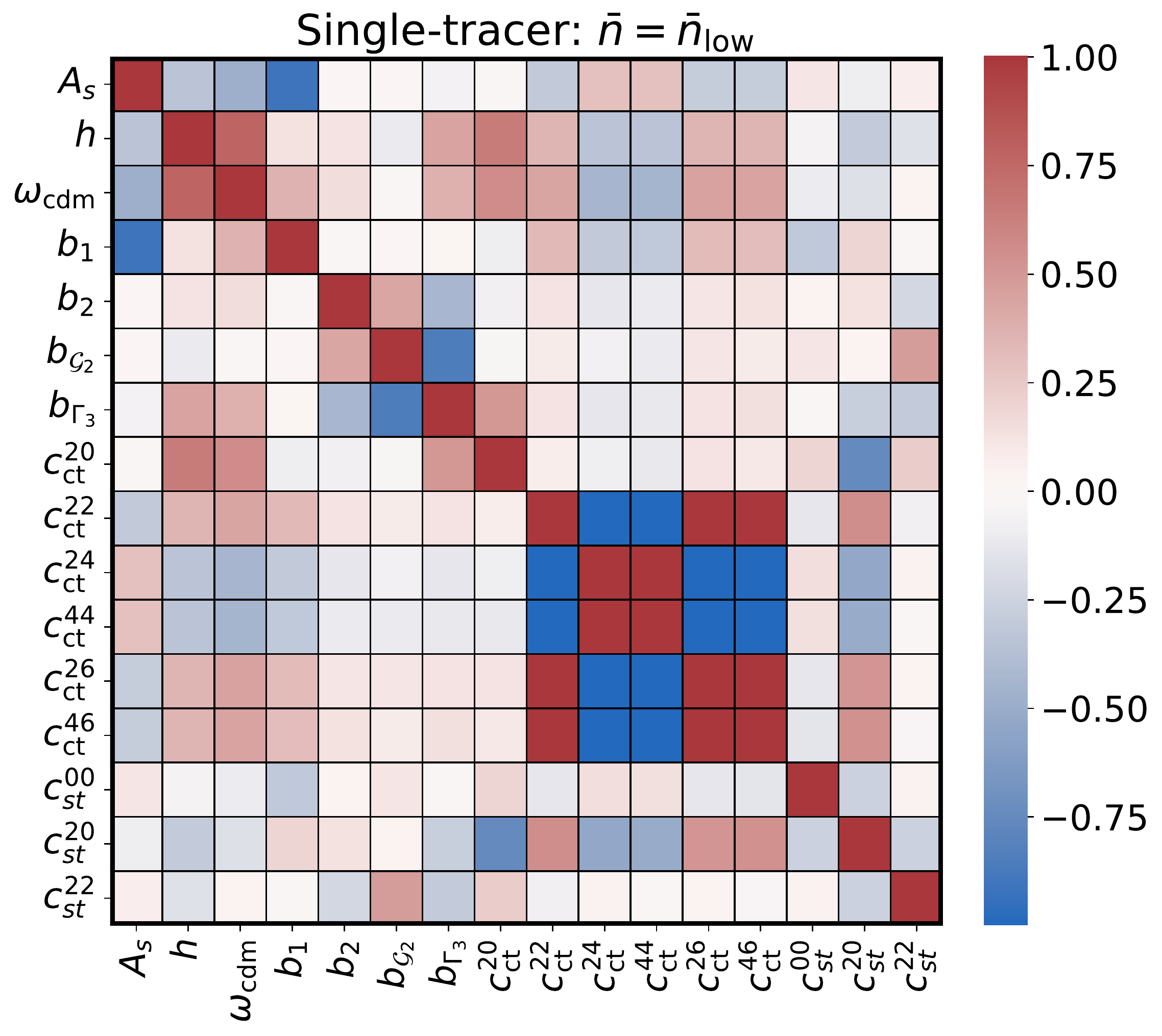}
    \includegraphics[width = 0.49\textwidth]{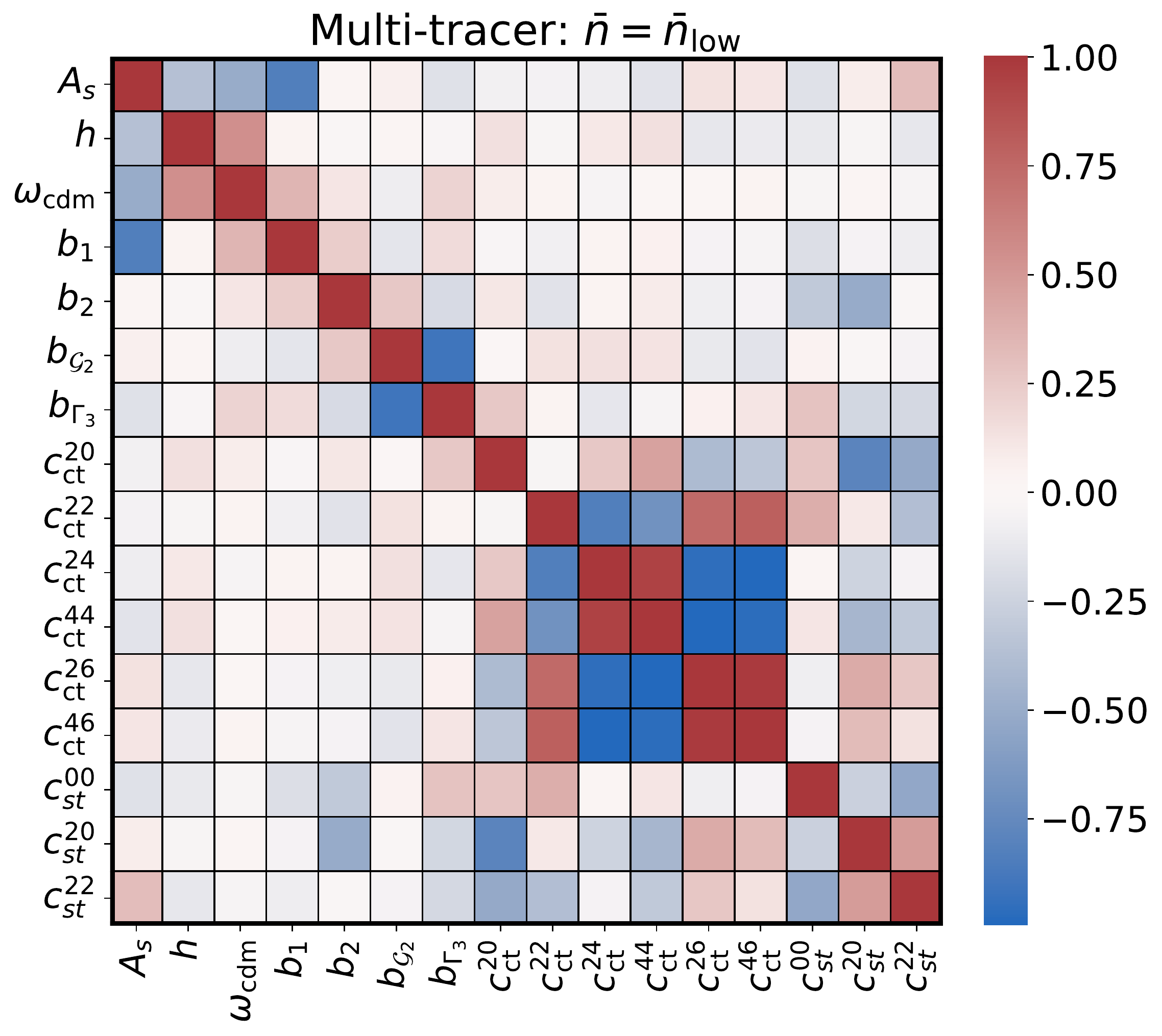}
    \caption{Correlation matrix between cosmological parameters, bias, stochastic coefficients and counter-terms for ST (left) and MT (right) for $k_{\rm max} = 0.14h/$Mpc. The top and bottom panels display values for the two shot noises considered in this work, $\bar{n}_{\rm high}$ (top) and $\bar{n}_{\rm low}$ (bottom). Although in the ST case there is a high correlation between $\{A_s,\,\omega_{\rm cdm},\,h\}$ and the other parameters, in the MT model part of  these correlations are broken. It indicates that the MT parameter basis is more orthogonal.} 
    \label{fig:correlation}
\end{figure}

To conclude, we show in Fig.~\ref{fig:correlation} the correlation matrix between the free parameter for ST (left) and MT (right). The matrix is more diagonal in the case of MT, especially in the cross-correlation between $c_{\rm ct}$'s and $\{A_s,\,\omega_{\rm cdm},h\}$, explaining part of the gains of MT compared to ST. Notice also that $b_1$ is less correlated with other parameters for MT. In addition, we highlight a strongly-correlated block (both for MT and ST) for the parameters $\{c_{\rm ct}^{22},\, c_{\rm ct}^{24},\, c_{\rm ct}^{44},\, c_{\rm ct}^{26},\, c_{\rm ct}^{46} \}$. The stochastic coefficients behave similarly for MT and ST. As we already mentioned, part of this gain can be attributed to the extra information coming from the cross spectrum, which constraints the product $b_\mathcal{O}^Ab_\mathcal{O}^B$. We also compare the correlations for $\bar{n}_{\rm high}$ (top) and $\bar{n}_{\rm low}$ (bottom). Notice that for the case with smaller shot noise ($\bar{n}_{\rm high}$) we find more correlations between the parameters if compared to the high shot noise scenario ($\bar{n}_{\rm low}$). For both values of $\bar{n}$ the correlation matrix for the effective tracer of MT is more diagonal. All that leads us to another important conclusion: {\it the MT basis of parameters is more orthogonal than the ST basis}, explaining part of the gains from MT. By breaking a population into distinct tracers, given the same data, the bias parameters and counter-term can encapsulate more information about the specific history and dynamics of each tracer. The combination of both populations into a single population blurs the data into an averaged tracer (see left panel of Fig.~\ref{fig:simulation}) and, consequently, less information is absorbed into each free coefficient. For instance, see the $b_1$ in the left panel of Fig.~\ref{fig:st_compared_single}: the two tracers present a different linear bias and the combination of both populations leads to an {\it average} tracer that erases part of the information gained when considering both populations separately.\footnote{Note that both MT and ST  present a linear bias that has an intermediate value of $b_1$ compared to considering $A$ and $B$ alone. However, in the MT case, the $b_1$ shown is from the {\it effective} tracer defined in Sec.~\ref{sec:combined_tracer} and not of the {\it average} tracer as ST. In practice, it means that while a single coefficient is fitted for ST, for MT we fit two coefficients using the auto and cross-spectra together, and later we take the average. We also highlight that MT's gains were observed despite the difference in the linear bias values for tracers $A$ and $B$ being relatively minor ($b_1^A \sim 0.8$ and $b_1^B \sim 1.5$). It indicates that the gain goes beyond the $b_1$ difference and linear scales.} MT is capable of making a clear estimate of the free coefficients for each specific sub-population, leading to a substantial improvement in the cosmological constraints. 

\section{Conclusions}
\label{sec:conclusion}

In this work, we extended the results of \cite{Mergulhao:2021kip} to redshift space, showing that the multi-tracer is still a relevant tool not only to beat the cosmic variance on linear scales but to break degeneracies between free parameters on the mildly non-linear regime. We also improved the previous analysis by considering more realistic galaxies generated by the state-of-the-art SHAMe method and a SFR tracer split.

In addition, we found that the FoG effects can be starker for different populations when doing the MT tracer splitting. We concluded that the higher-in-$\mu$ parameters in the EFT counter-terms are necessary to absorb those effects and extract unbiased cosmological parameters in an MT analysis. In addition, the relevance of those higher-in-$\mu$ terms indicate that it may be interesting to consider in the future the clustering wedges statistics from \cite{Kazin:2011xt,Sanchez:2013tga} for MT. Furthermore, we saw that the cross-stochastic terms did not deteriorate the MT improvement, in agreement with \cite{Mergulhao:2021kip}.

\begin{figure}
    \centering
    \includegraphics[width = \textwidth]{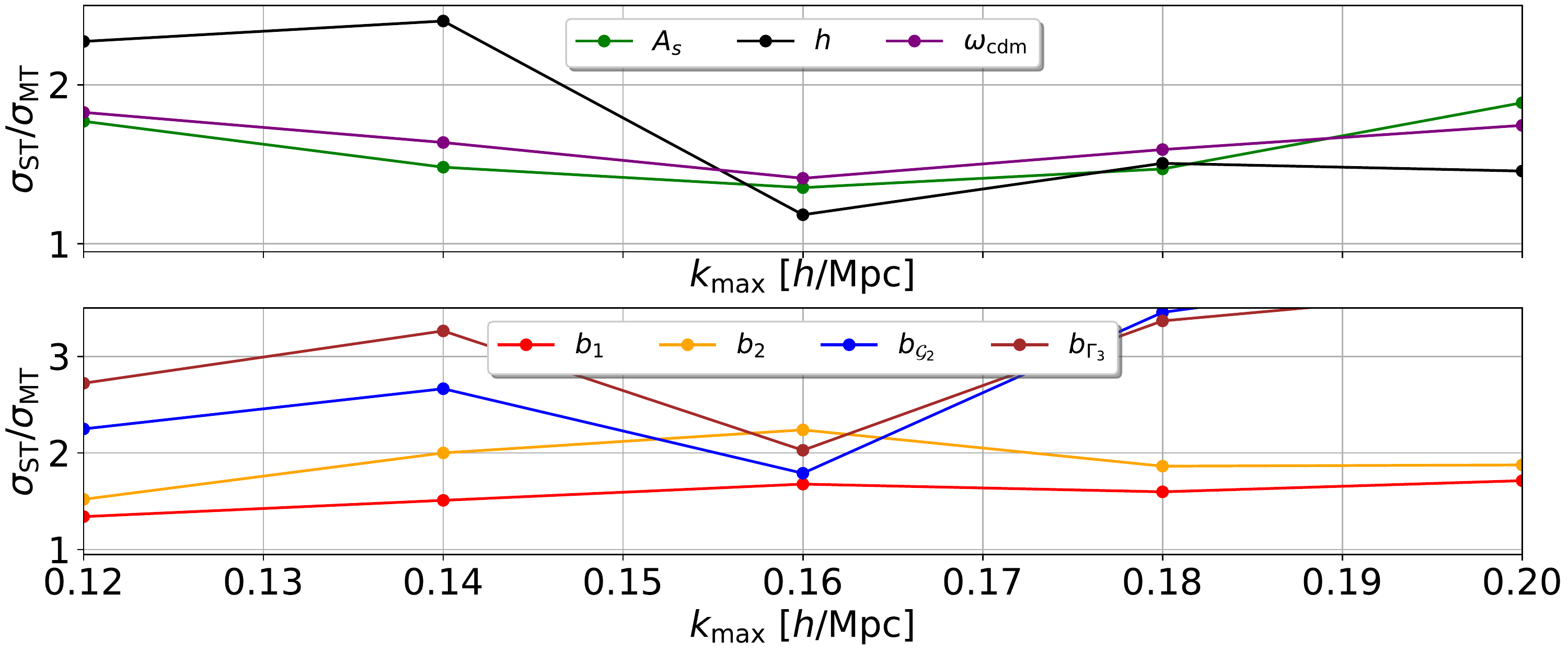}
    \caption{The multi-tracer (MT) improvements relative to the single-tracer (ST) scenario in a full-shape power-spectrum analysis. We display the size of the error bars of MT compared to ST for different values of $k_{\rm max}$. Our results show a consistent improvement of MT over ST for both cosmological (upper panel) and bias parameters (bottom panel). For the cosmological parameters, there is a consistent improvement $\sigma_{\rm ST}/\sigma_{\rm MT} \sim 1.5$. The improvements are even more striking for the bias parameters, especially for $b_{\Gamma_3}$ and $b_{\mathcal{G}_2}$, with MT error bars two or three times smaller than ST.
}
    \label{fig:my_label}
\end{figure}

We summarize in Fig.~\ref{fig:my_label} the main results of this work, comparing MT error bars relative to ST for all bias and cosmological parameters for $\bar{n} = \bar{n}_{\rm high}$ and different values of $k_{\rm max}$. We see that the error bars for MT are, on average, $50\%$ smaller for $A_s$, $h$, and $\omega_{\rm cdm}$. For the bias parameters, the error bars for MT are between 1.5 to 3 times better, with this improvement being smaller for $b_1$ and larger for $b_{\mathcal{G}_2}$ and $b_{\Gamma_3}$. Those improvements are consistent with the conclusions of the first paper \cite{Mergulhao:2021kip}, with the difference that we are now able to break the degeneracy between $A_s$ and $b_1$ in redshift space.

A common criticism to the EFTofLSS framework relies on a large number of free parameters present in the theory. This leads to a fundamental question: {\it when is it reasonable and justified to add new free parameters to an analysis?} The success of EFTofLSS when providing competitive cosmological constraints \cite{DAmico:2019fhj,Ivanov:2019hqk,Colas:2019ret, Philcox:2020vvt, Nishimichi:2020tvu, Semenaite:2021aen} shows that this criticism of EFT does {\it not} hold. The extra parameters added are {\it physical} and are exactly what guarantees the robustness of the theory by parametrizing the UV regime in a {\it self-controlled} manner. Those parameters are (relatively) well measured in data, which is one of the reasons why FS provides competitive constraints. In general, a $n$-tracer analysis will have about $n$-times more coefficients within the EFTofLSS framework. Again, we found that for the case considered in this work, those extra parameters do {\it not} jeopardize the constraints, but improve them. Despite sounding counter-intuitive to have a more constraining theory with more free parameters, the presence of the cross power spectrum between the tracers reduces the correlation between the EFT and the cosmological coefficients, which ultimately induce tighter constraints on both set of parameters. Those parameters are better determined for MT (Fig.~\ref{fig:fullmcmc}) and present fewer degeneracies (Fig.~\ref{fig:correlation}). In contrast, part of the information encoded in intermediate scales for each sub-tracer would be smoothed out when doing a single-tracer analysis (see the left panel of Fig.~\ref{fig:simulation}).

To conclude, MT provides an exciting perspective to maximize the information from galaxy surveys. There are several ways to split a sample of galaxies into sub-samples, and the MT gains might depend on how it is performed. In this work, we focused on a SFR split. It would be interesting in future work to explore other ways to split the tracers, such as considering the local environment density \cite{Paillas:2021oli}. Furthermore, this work is a fundamental step before applying MT to data, using simulations to test for the first time the EFTofLSS pipeline for MT in redshift space.

\acknowledgments
We are very thankful to Enea di Dio for providing his derivation for the covariance matrix for multi-tracer in redshift space. We also acknowledge Florian Beutler, John Peacock, and Pedro Carrilho for valuable comments, and Raul Abramo for his participation in the early stages of this work. HR acknowledges Mathias Garny and Fabian Schmidt for useful discussions and comments on the draft. HR is supported by the Deutsche Forschungsgemeinschaft under Germany's Excellence Strategy EXC 2094 `ORIGINS'. (No.\,390783311). TM is supported by the European Research Council (ERC) under the European Union’s Horizon 2020 research and innovation program (grant agreement 853291).

\bibliographystyle{JHEP}
\bibliography{main}

\providecommand{\href}[2]{#2}\begingroup\raggedright\begin{thebibliography}{10}

\bibitem{BOSS:2016wmc}
{\scshape BOSS} collaboration, S.~Alam et~al., \emph{{The clustering of
  galaxies in the completed SDSS-III Baryon Oscillation Spectroscopic Survey:
  cosmological analysis of the DR12 galaxy sample}},
  \href{http://dx.doi.org/10.1093/mnras/stx721}{\emph{Mon. Not. Roy. Astron.
  Soc.} {\bf 470} (2017) 2617--2652},
  [\href{http://arxiv.org/abs/1607.03155}{{\tt 1607.03155}}].

\bibitem{Abbetal}
{The Dark Energy Survey Collaboration}, \emph{{The Dark Energy Survey}},
  {\emph{ArXiv Astrophysics e-prints} (Oct., 2005) },
  [\href{http://arxiv.org/abs/astro-ph/0510346}{{\tt astro-ph/0510346}}].

\bibitem{SDSS-IV:2019txh}
{\scshape SDSS-IV} collaboration, R.~Ahumada et~al., \emph{{The 16th Data
  Release of the Sloan Digital Sky Surveys: First Release from the APOGEE-2
  Southern Survey and Full Release of eBOSS Spectra}},
  \href{http://dx.doi.org/10.3847/1538-4365/ab929e}{\emph{Astrophys. J. Suppl.}
  {\bf 249} (2020) 3}, [\href{http://arxiv.org/abs/1912.02905}{{\tt
  1912.02905}}].

\bibitem{DESI:2016fyo}
{\scshape DESI} collaboration, A.~Aghamousa et~al., \emph{{The DESI Experiment
  Part I: Science,Targeting, and Survey Design}},
  \href{http://arxiv.org/abs/1611.00036}{{\tt 1611.00036}}.

\bibitem{Amendola:2012ys}
{\scshape Euclid Theory Working Group} collaboration, L.~Amendola et~al.,
  \emph{{Cosmology and fundamental physics with the Euclid satellite}},
  \href{http://dx.doi.org/10.12942/lrr-2013-6}{\emph{Living Rev. Rel.} {\bf 16}
  (2013) 6}, [\href{http://arxiv.org/abs/1206.1225}{{\tt 1206.1225}}].

\bibitem{Ivezic:2008fe}
{\scshape LSST} collaboration, Z.~Ivezic, J.~A. Tyson, R.~Allsman, J.~Andrew
  and R.~Angel, \emph{{LSST: from Science Drivers to Reference Design and
  Anticipated Data Products}},  \href{http://arxiv.org/abs/0805.2366}{{\tt
  0805.2366}}.

\bibitem{Schoneberg:2021qvd}
N.~Sch\"oneberg, G.~Franco~Abell\'an, A.~P\'erez~S\'anchez, S.~J. Witte,
  V.~Poulin and J.~Lesgourgues, \emph{{The H0 Olympics: A fair ranking of
  proposed models}},
  \href{http://dx.doi.org/10.1016/j.physrep.2022.07.001}{\emph{Phys. Rept.}
  {\bf 984} (2022) 1--55}, [\href{http://arxiv.org/abs/2107.10291}{{\tt
  2107.10291}}].

\bibitem{Baumann:2010tm}
D.~Baumann, A.~Nicolis, L.~Senatore and M.~Zaldarriaga, \emph{{Cosmological
  Non-Linearities as an Effective Fluid}},
  \href{http://dx.doi.org/10.1088/1475-7516/2012/07/051}{\emph{JCAP} {\bf 07}
  (2012) 051}, [\href{http://arxiv.org/abs/1004.2488}{{\tt 1004.2488}}].

\bibitem{Carrasco:2012cv}
J.~J.~M. Carrasco, M.~P. Hertzberg and L.~Senatore, \emph{{The Effective Field
  Theory of Cosmological Large Scale Structures}},
  \href{http://dx.doi.org/10.1007/JHEP09(2012)082}{\emph{JHEP} {\bf 09} (2012)
  082}, [\href{http://arxiv.org/abs/1206.2926}{{\tt 1206.2926}}].

\bibitem{Carrasco:2013mua}
J.~J.~M. Carrasco, S.~Foreman, D.~Green and L.~Senatore, \emph{{The Effective
  Field Theory of Large Scale Structures at Two Loops}},
  \href{http://dx.doi.org/10.1088/1475-7516/2014/07/057}{\emph{JCAP} {\bf 07}
  (2014) 057}, [\href{http://arxiv.org/abs/1310.0464}{{\tt 1310.0464}}].

\bibitem{Konstandin:2019bay}
T.~Konstandin, R.~A. Porto and H.~Rubira, \emph{{The Effective Field Theory of
  Large Scale Structure at Three Loops}},
  \href{http://dx.doi.org/10.1088/1475-7516/2019/11/027}{\emph{JCAP} {\bf 11}
  (2019) 027}, [\href{http://arxiv.org/abs/1906.00997}{{\tt 1906.00997}}].

\bibitem{Angulo:2015eqa}
R.~Angulo, M.~Fasiello, L.~Senatore and Z.~Vlah, \emph{{On the Statistics of
  Biased Tracers in the Effective Field Theory of Large Scale Structures}},
  \href{http://dx.doi.org/10.1088/1475-7516/2015/9/029}{\emph{JCAP} {\bf 09}
  (2015) 029}, [\href{http://arxiv.org/abs/1503.08826}{{\tt 1503.08826}}].

\bibitem{Baldauf:2021zlt}
T.~Baldauf, M.~Garny, P.~Taule and T.~Steele, \emph{{Two-loop bispectrum of
  large-scale structure}},
  \href{http://dx.doi.org/10.1103/PhysRevD.104.123551}{\emph{Phys. Rev. D} {\bf
  104} (2021) 123551}, [\href{http://arxiv.org/abs/2110.13930}{{\tt
  2110.13930}}].

\bibitem{Assassi2014}
V.~Assassi, D.~Baumann, D.~Green and M.~Zaldarriaga, \emph{{Renormalized Halo
  Bias}}, \href{http://dx.doi.org/10.1088/1475-7516/2014/08/056}{\emph{JCAP}
  {\bf 08} (2014) 056}, [\href{http://arxiv.org/abs/1402.5916}{{\tt
  1402.5916}}].

\bibitem{Desjacques2016}
V.~Desjacques, D.~Jeong and F.~Schmidt, \emph{{Large-Scale Galaxy Bias}},
  \href{http://dx.doi.org/10.1016/j.physrep.2017.12.002}{\emph{Phys. Rep.} {\bf
  733} (nov, 2016) 1--193}, [\href{http://arxiv.org/abs/1611.09787}{{\tt
  1611.09787}}].

\bibitem{Sanchez:2013tga}
A.~G. Sanchez et~al., \emph{{The clustering of galaxies in the SDSS-III Baryon
  Oscillation Spectroscopic Survey: cosmological implications of the full shape
  of the clustering wedges in the data release 10 and 11 galaxy samples}},
  \href{http://dx.doi.org/10.1093/mnras/stu342}{\emph{Mon. Not. Roy. Astron.
  Soc.} {\bf 440} (2014) 2692--2713},
  [\href{http://arxiv.org/abs/1312.4854}{{\tt 1312.4854}}].

\bibitem{DAmico:2019fhj}
G.~D'Amico, J.~Gleyzes, N.~Kokron, K.~Markovic, L.~Senatore, P.~Zhang et~al.,
  \emph{{The Cosmological Analysis of the SDSS/BOSS data from the Effective
  Field Theory of Large-Scale Structure}},
  \href{http://dx.doi.org/10.1088/1475-7516/2020/05/005}{\emph{JCAP} {\bf 05}
  (2020) 005}, [\href{http://arxiv.org/abs/1909.05271}{{\tt 1909.05271}}].

\bibitem{Ivanov:2019hqk}
M.~M. Ivanov, M.~Simonovi\'c and M.~Zaldarriaga, \emph{{Cosmological Parameters
  and Neutrino Masses from the Final Planck and Full-Shape BOSS Data}},
  \href{http://dx.doi.org/10.1103/PhysRevD.101.083504}{\emph{Phys. Rev. D} {\bf
  101} (2020) 083504}, [\href{http://arxiv.org/abs/1912.08208}{{\tt
  1912.08208}}].

\bibitem{Colas:2019ret}
T.~Colas, G.~D'amico, L.~Senatore, P.~Zhang and F.~Beutler, \emph{{Efficient
  Cosmological Analysis of the SDSS/BOSS data from the Effective Field Theory
  of Large-Scale Structure}},
  \href{http://dx.doi.org/10.1088/1475-7516/2020/06/001}{\emph{JCAP} {\bf 06}
  (2020) 001}, [\href{http://arxiv.org/abs/1909.07951}{{\tt 1909.07951}}].

\bibitem{Philcox:2020vvt}
O.~H. Philcox, M.~M. Ivanov, M.~Simonovi\'c and M.~Zaldarriaga,
  \emph{{Combining Full-Shape and BAO Analyses of Galaxy Power Spectra: A 1.6\%
  CMB-independent constraint on H0}},
  \href{http://dx.doi.org/10.1088/1475-7516/2020/05/032}{\emph{JCAP} {\bf 05}
  (2020) 032}, [\href{http://arxiv.org/abs/2002.04035}{{\tt 2002.04035}}].

\bibitem{Nishimichi:2020tvu}
T.~Nishimichi, G.~D'Amico, M.~M. Ivanov, L.~Senatore, M.~Simonovi\'c, M.~Takada
  et~al., \emph{{Blinded challenge for precision cosmology with large-scale
  structure: results from effective field theory for the redshift-space galaxy
  power spectrum}},
  \href{http://dx.doi.org/10.1103/PhysRevD.102.123541}{\emph{Phys. Rev. D} {\bf
  102} (2020) 123541}, [\href{http://arxiv.org/abs/2003.08277}{{\tt
  2003.08277}}].

\bibitem{Semenaite:2021aen}
A.~Semenaite et~al., \emph{{Cosmological implications of the full shape of
  anisotropic clustering measurements in BOSS and eBOSS}},
  \href{http://dx.doi.org/10.1093/mnras/stac829}{\emph{Mon. Not. Roy. Astron.
  Soc.} {\bf 512} (2022) 5657--5670},
  [\href{http://arxiv.org/abs/2111.03156}{{\tt 2111.03156}}].

\bibitem{Ivanov:2020ril}
M.~M. Ivanov, E.~McDonough, J.~C. Hill, M.~Simonovi\'c, M.~W. Toomey,
  S.~Alexander et~al., \emph{{Constraining Early Dark Energy with Large-Scale
  Structure}}, \href{http://dx.doi.org/10.1103/PhysRevD.102.103502}{\emph{Phys.
  Rev. D} {\bf 102} (2020) 103502},
  [\href{http://arxiv.org/abs/2006.11235}{{\tt 2006.11235}}].

\bibitem{Lague:2021frh}
A.~Lagu\"e, J.~R. Bond, R.~Hlo\v{z}ek, K.~K. Rogers, D.~J.~E. Marsh and
  D.~Grin, \emph{{Constraining ultralight axions with galaxy surveys}},
  \href{http://dx.doi.org/10.1088/1475-7516/2022/01/049}{\emph{JCAP} {\bf 01}
  (2022) 049}, [\href{http://arxiv.org/abs/2104.07802}{{\tt 2104.07802}}].

\bibitem{Semenaite:2022unt}
A.~Semenaite, A.~G. S\'anchez, A.~Pezzotta, J.~Hou, A.~Eggemeier, M.~Crocce
  et~al., \emph{{Beyond \textendash{} \ensuremath{\Lambda}CDM constraints from
  the full shape clustering measurements from BOSS and eBOSS}},
  \href{http://dx.doi.org/10.1093/mnras/stad849}{\emph{Mon. Not. Roy. Astron.
  Soc.} {\bf 521} (2023) 5013--5025},
  [\href{http://arxiv.org/abs/2210.07304}{{\tt 2210.07304}}].

\bibitem{Rubira:2022xhb}
H.~Rubira, A.~Mazoun and M.~Garny, \emph{{Full-shape BOSS constraints on dark
  matter interacting with dark radiation and lifting the S8 tension}},
  \href{http://dx.doi.org/10.1088/1475-7516/2023/01/034}{\emph{JCAP} {\bf 01}
  (2023) 034}, [\href{http://arxiv.org/abs/2209.03974}{{\tt 2209.03974}}].

\bibitem{Piga:2022mge}
L.~Piga, M.~Marinucci, G.~D'Amico, M.~Pietroni, F.~Vernizzi and B.~S. Wright,
  \emph{{Constraints on modified gravity from the BOSS galaxy survey}},
  \href{http://dx.doi.org/10.1088/1475-7516/2023/04/038}{\emph{JCAP} {\bf 04}
  (2023) 038}, [\href{http://arxiv.org/abs/2211.12523}{{\tt 2211.12523}}].

\bibitem{Carrilho:2022mon}
P.~Carrilho, C.~Moretti and A.~Pourtsidou, \emph{{Cosmology with the EFTofLSS
  and BOSS: dark energy constraints and a note on priors}},
  \href{http://dx.doi.org/10.1088/1475-7516/2023/01/028}{\emph{JCAP} {\bf 01}
  (2023) 028}, [\href{http://arxiv.org/abs/2207.14784}{{\tt 2207.14784}}].

\bibitem{Simon:2022ftd}
T.~Simon, G.~Franco~Abell\'an, P.~Du, V.~Poulin and Y.~Tsai,
  \emph{{Constraining decaying dark matter with BOSS data and the effective
  field theory of large-scale structures}},
  \href{http://dx.doi.org/10.1103/PhysRevD.106.023516}{\emph{Phys. Rev. D} {\bf
  106} (2022) 023516}, [\href{http://arxiv.org/abs/2203.07440}{{\tt
  2203.07440}}].

\bibitem{Herold:2021ksg}
L.~Herold, E.~G.~M. Ferreira and E.~Komatsu, \emph{{New Constraint on Early
  Dark Energy from Planck and BOSS Data Using the Profile Likelihood}},
  \href{http://dx.doi.org/10.3847/2041-8213/ac63a3}{\emph{Astrophys. J. Lett.}
  {\bf 929} (2022) L16}, [\href{http://arxiv.org/abs/2112.12140}{{\tt
  2112.12140}}].

\bibitem{Seljak:2008xr}
U.~Seljak, \emph{{Extracting primordial non-gaussianity without cosmic
  variance}},
  \href{http://dx.doi.org/10.1103/PhysRevLett.102.021302}{\emph{Phys. Rev.
  Lett.} {\bf 102} (2009) 021302}, [\href{http://arxiv.org/abs/0807.1770}{{\tt
  0807.1770}}].

\bibitem{McDonald:2008sh}
P.~McDonald and U.~Seljak, \emph{{How to measure redshift-space distortions
  without sample variance}},
  \href{http://dx.doi.org/10.1088/1475-7516/2009/10/007}{\emph{JCAP} {\bf 10}
  (2009) 007}, [\href{http://arxiv.org/abs/0810.0323}{{\tt 0810.0323}}].

\bibitem{Abramo2013}
L.~R. Abramo and K.~E. Leonard, \emph{{Why multi-tracer surveys beat cosmic
  variance}}, \href{http://dx.doi.org/10.1093/mnras/stt465}{\emph{Mon. Not.
  Roy. Astron. Soc.} {\bf 432} (2013) 318},
  [\href{http://arxiv.org/abs/1302.5444}{{\tt 1302.5444}}].

\bibitem{Abramo:2015iga}
L.~R. Abramo, L.~F. Secco and A.~Loureiro, \emph{{Fourier analysis of
  multitracer cosmological surveys}},
  \href{http://dx.doi.org/10.1093/mnras/stv2588}{\emph{Mon. Not. Roy. Astron.
  Soc.} {\bf 455} (2016) 3871--3889},
  [\href{http://arxiv.org/abs/1505.04106}{{\tt 1505.04106}}].

\bibitem{Abramo:2022qir}
L.~R. Abramo, J.~a.~V. Dinarte~Ferri, I.~L. Tashiro and A.~Loureiro,
  \emph{{Fisher matrix for the angular power spectrum of multi-tracer galaxy
  surveys}}, \href{http://dx.doi.org/10.1088/1475-7516/2022/08/073}{\emph{JCAP}
  {\bf 08} (2022) 073}, [\href{http://arxiv.org/abs/2204.05057}{{\tt
  2204.05057}}].

\bibitem{Barreira:2023rxn}
A.~Barreira and E.~Krause, \emph{{Towards optimal and robust $f_{\rm NL}$
  constraints with multi-tracer analyses}},
  \href{http://arxiv.org/abs/2302.09066}{{\tt 2302.09066}}.

\bibitem{Karagiannis:2023lsj}
D.~Karagiannis, R.~Maartens, J.~Fonseca, S.~Camera and C.~Clarkson,
  \emph{{Multi-tracer power spectra and bispectra: Formalism}},
  \href{http://arxiv.org/abs/2305.04028}{{\tt 2305.04028}}.

\bibitem{Blake:2013nif}
C.~Blake et~al., \emph{{Galaxy And Mass Assembly (GAMA): improved cosmic growth
  measurements using multiple tracers of large-scale structure}},
  \href{http://dx.doi.org/10.1093/mnras/stt1791}{\emph{Mon. Not. Roy. Astron.
  Soc.} {\bf 436} (2013) 3089}, [\href{http://arxiv.org/abs/1309.5556}{{\tt
  1309.5556}}].

\bibitem{Ross:2013vla}
A.~J. Ross et~al., \emph{{The Clustering of Galaxies in the SDSS-III DR10
  Baryon Oscillation Spectroscopic Survey: No Detectable Colour Dependence of
  Distance Scale or Growth Rate Measurements}},
  \href{http://dx.doi.org/10.1093/mnras/stt1895}{\emph{Mon. Not. Roy. Astron.
  Soc.} {\bf 437} (2014) 1109--1126},
  [\href{http://arxiv.org/abs/1310.1106}{{\tt 1310.1106}}].

\bibitem{Beutler:2015tla}
F.~Beutler, C.~Blake, J.~Koda, F.~Marin, H.-J. Seo, A.~J. Cuesta et~al.,
  \emph{{The BOSS\textendash{}WiggleZ overlap region \textendash{} I. Baryon
  acoustic oscillations}},
  \href{http://dx.doi.org/10.1093/mnras/stv1943}{\emph{Mon. Not. Roy. Astron.
  Soc.} {\bf 455} (2016) 3230--3248},
  [\href{http://arxiv.org/abs/1506.03900}{{\tt 1506.03900}}].

\bibitem{Marin:2015ula}
F.~A. Mar\'\i{}n, F.~Beutler, C.~Blake, J.~Koda, E.~Kazin and D.~P. Schneider,
  \emph{{The BOSS\textendash{}WiggleZ overlap region \textendash{} II.
  Dependence of cosmic growth on galaxy type}},
  \href{http://dx.doi.org/10.1093/mnras/stv2502}{\emph{Mon. Not. Roy. Astron.
  Soc.} {\bf 455} (2016) 4046--4056},
  [\href{http://arxiv.org/abs/1506.03901}{{\tt 1506.03901}}].

\bibitem{Zhang:2021uyp}
P.~Zhang and Y.~Cai, \emph{{BOSS full-shape analysis from the EFTofLSS with
  exact time dependence}},
  \href{http://dx.doi.org/10.1088/1475-7516/2022/01/031}{\emph{JCAP} {\bf 01}
  (2022) 031}, [\href{http://arxiv.org/abs/2111.05739}{{\tt 2111.05739}}].

\bibitem{Sullivan:2023qjr}
J.~M. Sullivan, T.~Prijon and U.~Seljak, \emph{{Learning to Concentrate:
  Multi-tracer Forecasts on Local Primordial Non-Gaussianity with
  Machine-Learned Bias}},  \href{http://arxiv.org/abs/2303.08901}{{\tt
  2303.08901}}.

\bibitem{Wang:2020tje}
Y.~Wang et~al., \emph{{The clustering of the SDSS-IV extended Baryon
  Oscillation Spectroscopic Survey DR16 luminous red galaxy and emission line
  galaxy samples: cosmic distance and structure growth measurements using
  multiple tracers in configuration space}},
  \href{http://dx.doi.org/10.1093/mnras/staa2593}{\emph{Mon. Not. Roy. Astron.
  Soc.} {\bf 498} (2020) 3470--3483},
  [\href{http://arxiv.org/abs/2007.09010}{{\tt 2007.09010}}].

\bibitem{Zhao:2020tis}
G.-B. Zhao et~al., \emph{{The completed SDSS-IV extended Baryon Oscillation
  Spectroscopic Survey: a multitracer analysis in Fourier space for measuring
  the cosmic structure growth and expansion rate}},
  \href{http://dx.doi.org/10.1093/mnras/stab849}{\emph{Mon. Not. Roy. Astron.
  Soc.} {\bf 504} (2021) 33--52}, [\href{http://arxiv.org/abs/2007.09011}{{\tt
  2007.09011}}].

\bibitem{Zhao:2021ahg}
C.~Zhao et~al., \emph{{The completed SDSS-IV extended Baryon Oscillation
  Spectroscopic Survey: cosmological implications from multitracer BAO analysis
  with galaxies and voids}},
  \href{http://dx.doi.org/10.1093/mnras/stac390}{\emph{Mon. Not. Roy. Astron.
  Soc.} {\bf 511} (2022) 5492--5524},
  [\href{http://arxiv.org/abs/2110.03824}{{\tt 2110.03824}}].

\bibitem{Mergulhao:2021kip}
T.~Mergulh\~ao, H.~Rubira, R.~Voivodic and L.~R. Abramo, \emph{{The effective
  field theory of large-scale structure and multi-tracer}},
  \href{http://dx.doi.org/10.1088/1475-7516/2022/04/021}{\emph{JCAP} {\bf 04}
  (2022) 021}, [\href{http://arxiv.org/abs/2108.11363}{{\tt 2108.11363}}].

\bibitem{Angulo:2020vky}
R.~E. Angulo, M.~Zennaro, S.~Contreras, G.~Aric\`o, M.~Pellejero-Iba\~nez and
  J.~St\"ucker, \emph{{The BACCO simulation project: exploiting the full power
  of large-scale structure for cosmology}},
  \href{http://dx.doi.org/10.1093/mnras/stab2018}{\emph{Mon. Not. Roy. Astron.
  Soc.} {\bf 507} (2021) 5869--5881},
  [\href{http://arxiv.org/abs/2004.06245}{{\tt 2004.06245}}].

\bibitem{Zennaro:2021bwy}
M.~Zennaro, R.~E. Angulo, M.~Pellejero-Ib\'a\~nez, J.~St\"ucker, S.~Contreras
  and G.~Aric\`o, \emph{{The BACCO simulation project: biased tracers in real
  space}},  \href{http://arxiv.org/abs/2101.12187}{{\tt 2101.12187}}.

\bibitem{Contreras:2020him}
S.~Contreras, R.~Angulo and M.~Zennaro, \emph{{A flexible subhalo abundance
  matching model for galaxy clustering in redshift space}},
  \href{http://dx.doi.org/10.1093/mnras/stab2560}{\emph{Mon. Not. Roy. Astron.
  Soc.} {\bf 508} (2021) 175--189},
  [\href{http://arxiv.org/abs/2012.06596}{{\tt 2012.06596}}].

\bibitem{SFRcolor}
R.~{Feldmann}, E.~{Quataert}, P.~F. {Hopkins}, C.-A. {Faucher-Gigu{\`e}re} and
  D.~{Kere{\v{s}}}, \emph{{Colours, star formation rates and environments of
  star-forming and quiescent galaxies at the cosmic noon}},
  \href{http://dx.doi.org/10.1093/mnras/stx1120}{\emph{\mnras} {\bf 470}
  (Sept., 2017) 1050--1072}, [\href{http://arxiv.org/abs/1610.02411}{{\tt
  1610.02411}}].

\bibitem{Lazeyras:2019dcx}
T.~Lazeyras and F.~Schmidt, \emph{{A robust measurement of the first
  higher-derivative bias of dark matter halos}},
  \href{http://dx.doi.org/10.1088/1475-7516/2019/11/041}{\emph{JCAP} {\bf 11}
  (2019) 041}, [\href{http://arxiv.org/abs/1904.11294}{{\tt 1904.11294}}].

\bibitem{chudaykin2020nonlinear}
A.~Chudaykin, M.~M. Ivanov, O.~H.~E. Philcox and M.~Simonovi\'c,
  \emph{{Nonlinear perturbation theory extension of the Boltzmann code CLASS}},
  \href{http://dx.doi.org/10.1103/PhysRevD.102.063533}{\emph{Phys. Rev. D} {\bf
  102} (2020) 063533}, [\href{http://arxiv.org/abs/2004.10607}{{\tt
  2004.10607}}].

\bibitem{Perko2016}
A.~Perko, L.~Senatore, E.~Jennings and R.~H. Wechsler, \emph{{Biased Tracers in
  Redshift Space in the EFT of Large-Scale Structure}},
  \href{http://arxiv.org/abs/1610.09321}{{\tt 1610.09321}}.

\bibitem{Simon:2022lde}
T.~Simon, P.~Zhang, V.~Poulin and T.~L. Smith, \emph{{On the consistency of
  effective field theory analyses of BOSS power spectrum}},
  \href{http://arxiv.org/abs/2208.05929}{{\tt 2208.05929}}.

\bibitem{Jackson:1971sky}
J.~C. Jackson, \emph{{Fingers of God: A critique of Rees' theory of primoridal
  gravitational radiation}},
  \href{http://dx.doi.org/10.1093/mnras/156.1.1P}{\emph{Mon. Not. Roy. Astron.
  Soc.} {\bf 156} (1972) 1P--5P}, [\href{http://arxiv.org/abs/0810.3908}{{\tt
  0810.3908}}].

\bibitem{Ivanov2019}
M.~M. Ivanov, M.~Simonovi\'c and M.~Zaldarriaga, \emph{{Cosmological Parameters
  from the BOSS Galaxy Power Spectrum}},
  \href{http://dx.doi.org/10.1088/1475-7516/2020/05/042}{\emph{JCAP} {\bf 05}
  (2020) 042}, [\href{http://arxiv.org/abs/1909.05277}{{\tt 1909.05277}}].

\bibitem{Senatore:2014via}
L.~Senatore and M.~Zaldarriaga, \emph{{The IR-resummed Effective Field Theory
  of Large Scale Structures}},
  \href{http://dx.doi.org/10.1088/1475-7516/2015/02/013}{\emph{JCAP} {\bf 02}
  (2015) 013}, [\href{http://arxiv.org/abs/1404.5954}{{\tt 1404.5954}}].

\bibitem{Baldauf:2013hka}
T.~Baldauf, U.~Seljak, R.~E. Smith, N.~Hamaus and V.~Desjacques, \emph{{Halo
  stochasticity from exclusion and nonlinear clustering}},
  \href{http://dx.doi.org/10.1103/PhysRevD.88.083507}{\emph{Phys. Rev. D} {\bf
  88} (2013) 083507}, [\href{http://arxiv.org/abs/1305.2917}{{\tt 1305.2917}}].

\bibitem{2016MNRAS.462L...1A}
R.~E. {Angulo} and A.~{Pontzen}, \emph{{Cosmological N-body simulations with
  suppressed variance}},
  \href{http://dx.doi.org/10.1093/mnrasl/slw098}{\emph{\mnras} {\bf 462} (Oct.,
  2016) L1--L5}, [\href{http://arxiv.org/abs/1603.05253}{{\tt 1603.05253}}].

\bibitem{Maion:2022yjo}
F.~Maion, R.~E. Angulo and M.~Zennaro, \emph{{Statistics of biased tracers in
  variance-suppressed simulations}},
  \href{http://dx.doi.org/10.1088/1475-7516/2022/10/036}{\emph{JCAP} {\bf 10}
  (2022) 036}, [\href{http://arxiv.org/abs/2204.03868}{{\tt 2204.03868}}].

\bibitem{Moster_2018}
B.~P. Moster, T.~Naab and S.~D.~M. White, \emph{emerge {\textendash} an
  empirical model for the formation of galaxies since z~$\sim$~10},
  \href{http://dx.doi.org/10.1093/mnras/sty655}{\emph{Monthly Notices of the
  Royal Astronomical Society} {\bf 477} (mar, 2018) 1822--1852}.

\bibitem{Madgwick:2003bd}
D.~S. Madgwick et~al., \emph{{The 2dF galaxy redshift survey: Galaxy clustering
  per spectral type}},
  \href{http://dx.doi.org/10.1046/j.1365-8711.2003.06861.x}{\emph{Mon. Not.
  Roy. Astron. Soc.} {\bf 344} (2003) 847},
  [\href{http://arxiv.org/abs/astro-ph/0303668}{{\tt astro-ph/0303668}}].

\bibitem{Coil:2007jp}
A.~L. Coil et~al., \emph{{The DEEP2 Galaxy Redshift Survey: Color and
  luminosity dependence of galaxy clustering at z similar to 1}},
  \href{http://dx.doi.org/10.1086/523639}{\emph{Astrophys. J.} {\bf 672} (2008)
  153--176}, [\href{http://arxiv.org/abs/0708.0004}{{\tt 0708.0004}}].

\bibitem{Hang:2022zyb}
Q.~Hang, J.~A. Peacock, S.~Alam, Y.-C. Cai, K.~Kraljic, M.~van Daalen et~al.,
  \emph{{Galaxy and Mass Assembly (GAMA): probing galaxy-group correlations in
  redshift space with the halo streaming model}},
  \href{http://dx.doi.org/10.1093/mnras/stac2569}{\emph{Mon. Not. Roy. Astron.
  Soc.} {\bf 517} (2022) 374--392},
  [\href{http://arxiv.org/abs/2206.05065}{{\tt 2206.05065}}].

\bibitem{DiDioCovariance}
E.~Di~Dio, \emph{{{\rm et al.}, In preparation}}, .

\bibitem{Kaiser:1987qv}
N.~Kaiser, \emph{{Clustering in real space and in redshift space}}, {\emph{Mon.
  Not. Roy. Astron. Soc.} {\bf 227} (1987) 1--27}.

\bibitem{Sugiyama:2019ike}
N.~S. Sugiyama, S.~Saito, F.~Beutler and H.-J. Seo, \emph{{Perturbation theory
  approach to predict the covariance matrices of the galaxy power spectrum and
  bispectrum in redshift space}},
  \href{http://dx.doi.org/10.1093/mnras/staa1940}{\emph{Mon. Not. Roy. Astron.
  Soc.} {\bf 497} (2020) 1684--1711},
  [\href{http://arxiv.org/abs/1908.06234}{{\tt 1908.06234}}].

\bibitem{Wadekar:2019rdu}
D.~Wadekar and R.~Scoccimarro, \emph{{Galaxy power spectrum multipoles
  covariance in perturbation theory}},
  \href{http://dx.doi.org/10.1103/PhysRevD.102.123517}{\emph{Phys. Rev. D} {\bf
  102} (2020) 123517}, [\href{http://arxiv.org/abs/1910.02914}{{\tt
  1910.02914}}].

\bibitem{Blot_2019}
L.~Blot, M.~Crocce, E.~Sefusatti, M.~Lippich, A.~G. Sánchez, M.~Colavincenzo
  et~al., \emph{Comparing approximate methods for mock catalogues and
  covariance matrices ii: power spectrum multipoles},
  \href{http://dx.doi.org/10.1093/mnras/stz507}{\emph{Monthly Notices of the
  Royal Astronomical Society} {\bf 485} (Feb, 2019) 2806–2824}.

\bibitem{Foreman-Mackey:2012any}
D.~Foreman-Mackey, D.~W. Hogg, D.~Lang and J.~Goodman, \emph{{emcee: The MCMC
  Hammer}}, \href{http://dx.doi.org/10.1086/670067}{\emph{Publ. Astron. Soc.
  Pac.} {\bf 125} (2013) 306--312}, [\href{http://arxiv.org/abs/1202.3665}{{\tt
  1202.3665}}].

\bibitem{karamanis2022pocomc}
M.~Karamanis, D.~Nabergoj, F.~Beutler, J.~A. Peacock and U.~Seljak,
  \emph{{pocoMC: A Python package for accelerated Bayesian inference in
  astronomy and cosmology}},
  \href{http://dx.doi.org/10.21105/joss.04634}{\emph{J. Open Source Softw.}
  {\bf 7} (2022) 4634}, [\href{http://arxiv.org/abs/2207.05660}{{\tt
  2207.05660}}].

\bibitem{karamanis2022accelerating}
M.~Karamanis, F.~Beutler, J.~A. Peacock, D.~Nabergoj and U.~Seljak,
  \emph{{Accelerating astronomical and cosmological inference with
  preconditioned Monte Carlo}},
  \href{http://dx.doi.org/10.1093/mnras/stac2272}{\emph{Mon. Not. Roy. Astron.
  Soc.} {\bf 516} (2022) 1644--1653},
  [\href{http://arxiv.org/abs/2207.05652}{{\tt 2207.05652}}].

\bibitem{Gelman:1992zz}
A.~Gelman and D.~B. Rubin, \emph{{Inference from Iterative Simulation Using
  Multiple Sequences}},
  \href{http://dx.doi.org/10.1214/ss/1177011136}{\emph{Statist. Sci.} {\bf 7}
  (1992) 457--472}.

\bibitem{Blas:2011rf}
D.~Blas, J.~Lesgourgues and T.~Tram, \emph{{The Cosmic Linear Anisotropy
  Solving System (CLASS) II: Approximation schemes}},
  \href{http://dx.doi.org/10.1088/1475-7516/2011/07/034}{\emph{JCAP} {\bf 07}
  (2011) 034}, [\href{http://arxiv.org/abs/1104.2933}{{\tt 1104.2933}}].

\bibitem{Ivanov:2019pdj}
M.~M. Ivanov, M.~Simonovi\'c and M.~Zaldarriaga, \emph{{Cosmological Parameters
  from the BOSS Galaxy Power Spectrum}},
  \href{http://dx.doi.org/10.1088/1475-7516/2020/05/042}{\emph{JCAP} {\bf 05}
  (2020) 042}, [\href{http://arxiv.org/abs/1909.05277}{{\tt 1909.05277}}].

\bibitem{Kazin:2011xt}
E.~A. Kazin, A.~G. Sanchez and M.~R. Blanton, \emph{{Improving measurements of
  H(z) and Da(z) by analyzing clustering anisotropies}},
  \href{http://dx.doi.org/10.1111/j.1365-2966.2011.19962.x}{\emph{Mon. Not.
  Roy. Astron. Soc.} {\bf 419} (2012) 3223--3243},
  [\href{http://arxiv.org/abs/1105.2037}{{\tt 1105.2037}}].

\bibitem{Paillas:2021oli}
E.~Paillas, Y.-C. Cai, N.~Padilla and A.~G. S\'anchez, \emph{{Redshift-space
  distortions with split densities}},
  \href{http://dx.doi.org/10.1093/mnras/stab1654}{\emph{Mon. Not. Roy. Astron.
  Soc.} {\bf 505} (2021) 5731--5752},
  [\href{http://arxiv.org/abs/2101.09854}{{\tt 2101.09854}}].

\end{thebibliography}\endgroup

\end{document}